%% file: Reid8.tex
\documentclass[preprint]{aastex} 

\slugcomment{Accepted: 26-03-2004}
\shorttitle{2MASS Ultracool Dwarfs}
\shortauthors{Reid et al.}

\begin{document}
\def\pedant{Mart{\'{\i}}n }
\def\etal{{\sl et al.}}
\def\etall{{\sl et al. }}
\def\pma{$\arcsec$~yr$^{-1}$ }
\def\kms{km~s$^{-1}$ }
\def\msun{$M_{\odot}$}
\def\rsun{$R_{\odot}$}
\def\lsun{$L_{\odot}$}
\def\halpha{H$\alpha$}
\def\hbeta{H$\beta$}
\def\hgama{H$\gamma$}
\def\hdelta{H$\delta$}
\def\Teff{T$_{eff}$}
\def\logg{$log_g$}

\title{Meeting the Cool Neighbors VIII: A preliminary 20-parsec census from the NLTT catalogue}

\author{I. Neill Reid\altaffilmark{1}}
\affil{Space Telescope Science Institute, 3700 San Martin Drive, 
Baltimore, MD 
21218; and
Department of Physics and Astronomy, University of Pennsylvania, 209 
South 33rd 
Street,
Philadelphia, PA  19104; inr@stsci.edu}

\author{Kelle L. Cruz\altaffilmark{1}, Peter R. Allen\altaffilmark{1}, 
F. Mungall\altaffilmark{1}}
\affil{Department of Physics and Astronomy, University of Pennsylvania, 
209 South 33rd Street, Philadelphia, PA  19104}

\author{D. Kilkenny}
\affil{South African Astronomical Observatory, P.O. Box 9, Observatory 
7935, 
South Africa}

\author{James Liebert\altaffilmark{1}}
\affil{Department of Astronomy and Steward Observatory, 
University of Arizona, Tucson, AZ 85721}

\author{Suzanne L. Hawley, Oliver Fraser, Kevin Covey}
\affil{Department of Astronomy, University of Washington, Seattle, WA 
28195}

\author{Patrick Lowrance\altaffilmark{1}, J. Davy Kirkpatrick}
\affil{IPAC, California Institute of Technology, Pasadena, CA 91125}

\author {Adam J. Burgasser}
\affil{University of California, Los Angeles, }

\altaffiltext{1}{Visiting Astronomer, Kitt Peak National Observatory, 
NOAO, which is operated by AURA under cooperative agreement with the NSF.}

\begin{abstract}

Continuing our census of late-type dwarfs in the Solar Neighbourhood, 
we present BVRI photometry and optical spectroscopy of 800 mid-type M dwarfs
drawn from the NLTT proper motion catalogue. The targets are taken from both 
our own cross-referencing of the NLTT catalogue and the 2MASS Second Incremental
release, and from the revised NLTT compiled by Salim \& Gould (2003).
All are identified as nearby-star candidates based on their 
location in the (m$_r$, (m$_r$-K$_S$)) diagram. Three hundred 
stars discussed here
have previous astrometric, photometric or spectroscopic observations.
We present new BVRI photometry for 101 stars, together with low resolution
spectroscopy of a further 400 dwarfs. In total, we find that 241 stars are
within 20 parsecs of the Sun, while a further 70 lie within 1$\sigma$ of our
distance limit. \\
Combining the present results with 
previous analyses, we have quantitative observations for 1910 of the 1913 
candidates in our NLTT nearby-star samples. Eight hundred and fifteen of
those stars have distance estimates of 20 parsecs or less, including 312
additions to the local census. \\
With our NLTT follow-up observations essentially complete, we have
searched the literature for K and early-type M dwarfs within the 
sampling volume covered by the 2MASS Second Release.
Comparing the resultant 20-parsec census against predicted numbers, 
derived from the 8-parsec luminosity function, shows an overall deficit of
$\sim20\%$ for stellar systems and $\sim35\%$ for
individual stars. Almost all are likely to be fainter than
M$_J$=7, and at least half are probably as-yet undiscovered companions
of known nearby stars. Our results suggest that there are relatively
few missing {\sl systems} at the lowest luminosities, M$_J > 8.5$. 
We discuss possible means of identifying the missing stars.

\end{abstract}


\mbox{}

\section{Introduction}

It is by now a clich\'e (particularly after this set of papers) that
M dwarfs are the most populous members of the Galactic Disk. M subdwarfs
probably dominate the Galactic halo to a similar extent. It
is also now well established that  
the most recently-published local census, the Third Catalogue of Nearby Stars
(Gliese \& Jahreiss, 1991, CNS3), becomes incomplete for early-type 
dwarfs at a distance
of $\sim20-22$ parsecs, and is barely complete within 5 parsecs
for spectral types later than M6 (Reid, Gizis \& Hawley, 
2002 - PMSU4; Henry \etal, 2003). Thus, while Hipparcos (ESA, 1997) 
has given us a complete catalogue of over 1000 G dwarfs within 30 parsecs of 
the Sun, the statistical properties of dwarfs like Proxima Cen rest 
on observations of a volume-complete sample of a mere half-dozen stars.

We are using data from the 2-Micron All-Sky Survey (2MASS) (2MASS: Skrutskie
et al, 1997) to address this issue in a project carried out under the auspices
of the NASA/NSF NStars Initiative. Our goal is to compile a statistically
complete census of late-type dwarfs (spectral types late-K, M and L) 
within 20 parsecs of the Sun. With the publication of the eighth paper in
this series, we provide an assessment of progress toward that goal
and map out likely developments in the immediate future.

So far, we have employed two distinct, complementary 
techniques to identify unrecognised constituents of the immediate Solar 
Neighbourhood: first, we have used near-infrared photometry to
identify ultracool dwarfs (spectral types M7 and later) directly from 
the 2MASS database; second, we have searched for nearby early- and 
mid-type M dwarfs by cross-referencing stars in the
NLTT proper motion catalogue (Luyten, 1980) against the 2MASS data.
We have pursued both avenues of exploration in parallel.  
In both cases, our analysis to date is restricted to the 48\% of the sky
covered by the 2MASS Second Incremental Release (hereinafter, the 2M2nd), 
with only limited coverage of regions within 10 degrees of the Galactic Plane.

Ultracool dwarfs can be identified directly from 2MASS data alone 
since they have distinctively 
red near-infrared colours, with (J-K$_S$) increasing from a colour of $\sim$1.0
mag at spectral type M7.5 to $>1.8$ mag for late-type L dwarfs, until the onset
of significant methane absorption reverses this trend in L8 dwarfs 
at temperatures of
1300-1400K (Tsuji, 1964; Kirkpatrick \etal, 1999; Golimowski \etal, 2004). 
We have used the location of 2MASS point sources in both 
the (J, (J-K$_S$)) colour-magnitude
diagram and in the two-colour (J-H)/(H-K$_S$) to select candidates for
dwarfs within 20 parsecs of the Sun. 
The results from this thread of our project are published as follows:
\begin{itemize}
\item Paper IV ( Reid \etal, 2002) presents the initial result from the
ultracool survey, the identification of an M8 dwarf within 6 parsecs of the Sun.
\item Paper V (Cruz \etal, 2003) provides a thorough description of the
selection criteria used to compile the ultracool sample (the 2MU2
sample), and discusses optical spectroscopy of 298 candidates. The 2MU2 sample
is limited to $|b| > 10^o$, and specifically excludes areas near star-formation
regions.
\item Paper VI (Reid, 2003) represents a first attempt to extend 
photometric  ultracool surveys to the Galactic Plane, with only limited success.
\item Paper IX (Cruz \etal, 2004) presents observations of the remaining
2MU2 targets accessible to optical spectroscopy on 4-metre class telescopes, discusses
sample completeness and derives the luminosity function for spectral types M8 to L7.
\end{itemize}
Building on these results, we have used analysis of the 2MASS All-Sky data
Release to compile an all-sky high-latitude ($|b|>15^o$) sample of ultracool
dwarfs. Follow-up observations of that sample are almost complete, and 
the initial results, based on optical spectroscopy, will be presented in
Paper X in this series. 
In addition, we have obtained infrared spectroscopy of the faintest ultracool 
candidates from both samples, and those results will be presented in Paper XI.

Early- and mid-type M dwarfs do not possess the easily recogniseable near-infrared
colours of ultracool dwarfs, so additional information is required to segregate
the nearby stars among the 10$^8$ sources in the 2M2nd point-source catalogue.
Proper motion offers a well-tried method of identifying stars in the
vicinity of the Sun. The most extensive proper motion catalogues currently
available are those compiled by W. Luyten using photographic material,
primarily plates taken with the 48-inch Palomar Oschin Schmidt.
For a variety of reasons, principally the limited information offered
by the catalogued photometry, these resources have been little exploited. However, 
Monte-Carlo simulations based standard disk kinematics (PMSU4) show 
that between 88 and 90\% of stars
within 20 parsecs are expected to have motions $\mu > 0.18$ \pma, 
the proper motion limit of Luyten's NLTT survey.  
We have therefore cross-referenced
the NLTT catalogue against the 2MASS database, and searched for
nearby-star candidates using the optical-infrared
colours derived by combining Luyten's red magnitudes, m$_r$, with the 2MASS
JHK$_S$ photometry. The results from this second thread have appeared as
follows:
\begin{itemize}
\item Paper I (Reid \& Cruz, 2002) outlines the optical/near-infrared 
colour-magnitude criteria that have been used to identify candidate 
nearby stars. Based on those criteria, we selected two parent
samples of NLTT dwarfs, limiting the analysis to $|b| > 10^o$:
NLTT1, consisting of 1237 NLTT stars with 2M2nd counterparts
within 10 arcseconds of the predicted NLTT location; and 
stars not included in the NLTT1 sample, but which have 2M2nd sources within 
60 arcseconds of the predicted position. Literature data were compiled for 
469 stars, and distance estimates derived using (V-I), (I-J) and (V-K$_S$)
colour-magnitude relations. We place X of those stars within 20 parsecs of 
the Sun.
\item Paper II (Reid, Kilkenny \& Cruz, 2002) presents BVRI photometry of 180
southern NLTT1 stars and the corresponding photometric parallax estimates.
Y stars have formal distances estimates less than 20 parsecs.
\item Paper III (Cruz \& Reid, 2002) describes spectroscopic observations of
seventy NLTT dwarfs, and summarises the techniques adopted to
derive both spectral types and spectrophotometric parallax.
X stars fall within our 20 parsec distance limit.
\item Paper VII (Reid \etal, 2003) includes photometric and spectroscopic
observations of 453 dwarfs from the NLTT1 sample; we find that 111 of those
stars are likely to lie within 20 parsecs of the Sun. 
\item The present paper completes our NLTT survey, presenting photometric 
and spectroscopic observations and distance estimates for not only the
remaining stars in the NLTT1 sample, but also for candidate nearby stars
from two other NLTT samples, NLTT2 and NLTT3, where the latter is derived 
from Salim \& Gould's (2003) revised NLTT catalogue.
\end{itemize}
We do not propose to extend our NLTT-based survey to cover the full
celestial sphere by extending analysis to the 2MASS All-Sky data release -
with observations of half the sky, we have sufficient numbers for 
statistically-significant analyses of the intrinsic properties of
K- and M-type dwarfs. 
However, as discussed in more detail in the concluding sections
of this paper, we are exploring the potential for using other techniques to
identify nearby early- and mid-type M dwarfs that lie within the area covered 
by the 2M2nd data release, but are not in the NLTT catalogue.

Previous papers in the NLTT series have concentrated on follow-up
observations of the NLTT1 sample. 
23,795 of the 58,818 NLTT stars meet this criterion; distance
estimates for 1090 of the 1237 NLTT1 nearby-star candidates are given in 
Papers I-VII. The current paper completes observations of that sample, and
extends coverage to nearby-star candidates from two other 
NLTT-based samples. First, as described in Paper I, the NLTT1 stars 
exhibit distinct `holes' in their ($\alpha, \delta$) - notably near the North
Celestial Pole and the South Galactic Pole. Many of the 
unmatched NLTT stars in those holes have 2M2nd sources within 
60 arcseconds (see Figure 3 in Paper I), suggesting the presence of 
systematic offsets in the NLTT astrometry. 
The latter stars, together with NLTT/2MASS pairs lying at low
latitude ($|b| < 10^o$), constitute the NLTT2 sample.

Second, we have taken advantage of the independent analysis of
the NLTT catalogue by Salim \& Gould (2003 - SG03). As with our own
study, these authors match NLTT stars against the 2M2nd database, although
their analysis is limited to regions which overlap the POSS I survey
(effectively, $\delta > -33^o$). SG03 have 
gone to considerable length to correct typographical errors in 
the original catalogue, and they recover 97\% of the NLTT dwarfs brighter
than m$_r$=17 and 85\% to m$_r$=18. Coupled with their analysis of Hipparcos
and Tycho data for brighter stars (Gould \& Salim, 2003), the revised 
NLTT catalogue (rNLTT) includes data for 28379 stars with infrared
photometry from the 2M2nd database. Covering almost the same ground
as our own survey, the rNLTT provides both independent verification of 
our NLTT1 and NLTT2 samples, and a means of identifying new candidates.
We have therefore compiled a third list of candidate nearby stars, 
the NLTT3 sample, by applying our (m$_r$, (m$_r$-K$_S$)) selection
criteria to the rNLTT.

The present paper is organised in the following manner: 
Section 2 describes the NLTT2 sample in more detail, summarising 
previously-published photometric and spectroscopic data available for
those stars. Section 3 provides a similar function for the NLTT3 sample. 
New BVRI photometric observations of stars from both samples are 
presented in \S4, while \S5 presents optical spectroscopy of stars from
all three samples, NLTT1, NLTT2 and NLTT3. Combining all of these
results, we can derive reliable distance estimates for 1910 of the 1913 NLTT
nearby-star candidates.
Section 6 expands the discussion to include early-type M and K dwarfs, drawn
from the NLTT and other catalogues, and combines all of the datasets to 
give a 20-parsec census for the part of the sky covered by the 2M2nd. We
discuss the statistical properties of the resulting catalogue,
notably the luminosity function and the binary fraction, in \S7
and compare those properties
against similar data for the northern 8-parsec sample (Paper IV) and 
for the 25-parsec bright-star census compiled by PMSU4. Section 8
outlines possible methods for identifying additional stars within our
20-parsec limit, and the final section, \S9, presents our conclusions.

\section {The NLTT2 Sample}

\subsection {Defining the sample}

Approximately 34,000 NLTT stars either lack a 2M2nd source
within 10 arcseconds of the predicted position, or lie within 10 degrees of the
Galactic Plane. These stars are not included in the NLTT1 sample, and the 
overwhelming majority lie beyond the bounds of the 2M2nd. However,
expanding the search radius to 60 arcseconds for proper
motion stars with $|b| > 10^o$ results in matches between 5720 NLTT stars and 
25305 2MASS sources. We have also cross-referenced the two
catalogues at lower Galactic latitudes, but with the search radius 
limited to 10 arcseconds to minimise confusion. A total of 886 NLTT dwarfs are 
matched against 1951 2MASS sources. The final NLTT2 sample is derived by
combining these two datasets.

Most of the pairings generated by this exercise are obviously spurious - 
with a search radius of 60 arcseconds, many NLTT stars are matched
against six or more separate 2MASS sources, even at moderate Galactic latitudes. 
We have trimmed the list by applying the (m$_r$, (m$_r$-K$_S$)) 
colour-magnitude selection criteria outlined in Paper I. This reduces the
sample to 1468 NLTT/2MASS pairs with colours consistent with nearby M dwarfs,
but many of those pairings have incompatible near-infrared colours. 
Figure 1 plots the ((m$_r$-J)/(J-K$_S$)) and ((J-H)/(H-K$_S$))
two-colour diagrams for all 1468 candidates. The latter diagram also shows
the locii of main-sequence dwarfs and red giants, and we have outlined schematically the regions occupied by dwarfs later than spectral type 
$\approx$M3 ((H-K$_S) \ge 0.19$)). There are obviously many mismatched red
giants among the current candidates, which is not surprising given that this
dataset includes many NLTT dwarfs close to the Plane. Eliminating pairs
which have JHK$_S$ colours inconsistent with M or L dwarfs reduces the NLTT2
candidate list to 514 targets. 

Thirty-seven of the remaining 514 candidates fall in the L dwarf
segment of the JHK$_S$ diagram, although most lie suspiciously
close to the giant sequence. Unfortunately, all of these candidates
are found in implausible locations in the (m$_r$-J)/(J-K$_S$) diagram,
with most having implausibly blue (m$_r$-J) colours, so we have not
succeeded in finding an L dwarf in the NLTT catalogue. 
These are likely to be mismatches
to late-type red giants, lying on the redder edge of the giant sequence in the JHK$_S$ diagram. We have eliminated all 34 accordingly. Visual
inspection of 2MASS and POSS/UKST sky survey photographic images shows that
a further 111 pairings are mis-matches, either between components in
a common proper-motion binary or with a random field star. Thus, the 
final NLTT2 includes 369 candidate nearby M dwarfs. Of these, 
117, or approximately one-third, lie within 10 degrees of the Plane.

\subsection {Known nearby stars}

A substantial number of NLTT2 stars are already known as 
denizens of the immediate Solar Neighbourhood. One hundred and
nineteen stars are included in the CNS3, and most of these were
observed as part of the Palomar-Michigan State
University Spectroscopic survey (Reid, Hawley \& Gizis, 1995 - PMSU1; 
Hawley, Gizis \& Reid, 1996 - PMSU2).
Other stars have published spectroscopy, multicolour photometry or 
even trigonometric parallax measurements. Those measurements are sufficient 
to allow reliable distance estimation for 168 of the 369 NLTT2 candidate
nearby stars. 

Distances are based on the following sources:
65 stars were either observed directly by Hipparcos or are companions of
brighter stars with Hipparcos data, and have 
milliarcsecond-accuracy trigonometric
parallaxes. A further 48 stars have ground-based trigonometric parallaxes, including
forty-one with measurements accurate to ${\sigma_\pi \over \pi} < 0.15$. 
One hundred and twelve NLTT2 stars have published VRI photometry and
an additional 56 stars have BV photometry, 
allowing photometric parallaxes to be estimated using
the colour-magnitude relations given in Paper I.
Finally, 110 stars have published spectroscopy and spectroscopic parallax estimates.
Table 1 collects the available data and summarises the resulting 
distance estimates. Ninety-six of the 168 stars in this table have formal distance estimates less than 20 parsecs, while a further 12 stars lie within 1$\sigma$ of the 20-parsec limit.

\section {The NLTT3 Sample}

\subsection {Defining the sample}

We have selected nearby star candidates from the rNLTT (Salim \&
Gould, 2003) using the 
same (m$_r$, (m$_r$-K$_S$)) colour-magnitude criteria employed in
constructing the NLTT1 and NLTT2 samples. A total of 1908 stars
meet those criteria, including 1523 stars that are already included
in either the NLTT1 or NLTT2 samples. We have cross-referenced the latter
stars against our own results, and find no discrepancies - that is, both
our analysis and the rNLTT match the NLTT dwarfs against the same 2M2nd
sources. This includes several cases where that match is incorrect,
as discussed further below. The remaining 385 stars are the base
NLTT3 sample. A number of those candidates, however, have optical/near-infrared
colours that are inconsistent with those of late-type dwarfs.

The rNLTT supplements the red and photographic magnitudes given
in the original NLTT with photometry from a variety of sources, 
notably the USNOA catalogue (Monet \etal, 1998) and the Tycho
database (H{\o}g \etal, 2000). SG03 use those additional data to estimate 
V magnitudes for each source. Figure 2 plots colour-colour diagrams for all 1908 
colour-selected rNLTT stars. The majority have near-infrared
colours consistent the M dwarf sequence, but a number of sources are
unexpectedly blue in either (V-J) or (J-K$_S$) for the inferred (m$_r$-K$_S$).

We have identified 46 of the 1908 rNLTT nearby-star
candidates which meet at least one of the following criteria:
\begin{displaymath}
(m_r-K_S) \quad > \quad 14.44 (J-K_S) \quad - \quad 4.44
\end{displaymath}
\begin{displaymath}
(m_r-K_S) \quad > \quad 0.91 (V-J) \quad + \quad 3.09 
\end{displaymath}
These linear relations are plotted in Figure 2.
We have checked each outlier using both 2MASS images, provided by 
the Infrared Source Archive (IRSA), and digitised scans of the POSS/UKST 
sky survey plates maintained at Canadian Data Center, and our
conclusions are listed in Appendix I. In a number
of cases, the fault lies with either the NLTT photometry or the 2MASS
data (usually saturation in the 2M2nd) and, generally, those stars are
retained in our observing program. However, 
the majority are mis-identifications or merged images. In most cases, 
the mismatched stars are from common proper-motion binary systems, 
where the fainter optical component (often
a white dwarf) has been matched against 2MASS data for the primary, resulting 
in an apparently-red (m$_r$-K$_S$) colour. SG03 explicitly identified this
as a potential issue in using the rNLTT, and some of the same 
mis-matches were in our original NLTT1 sample. 

Overall, the rNLTT has a false identification rate of only $\sim1\%$
(SG03). However, our selection criteria, targeting sources 
with the reddest (m$_r$-K$_S$) colours, are ideally suited for turning up
mis-matched optical/near-infrared sources. 
We have therefore used the IRSA 2MASS images and photographic sky-survey 
data to verify that the base NLTT3 candidates are matched correctly.
Appendix II lists a further 63 entries from the rNLTT which either are mismatches
with the incorrect 2MASS source or have unreliable (m$_r$-K$_S$) colours
due to the close proximity of another star (usually a binary companion).
Combining appendices I and II, 78 stars in the base NLTT3 sample are eliminated,
giving a final target list of 307 nearby-star candidates from the rNLTT. 

\subsection {Known nearby stars}

Many NLTT3 stars have previous observations, and we can estimate reliable
distances in 151 cases, almost half the sample. The relevant data for 
those stars are given in Table 2. 
Ninety-seven stars have trigonometric parallax data, including 68 stars with 
either direct Hipparcos astrometry or Hipparcos measurements of bright companions.  
One hundred and forty-eight stars have at least V-band photometry, including eighty 
with V(R)I measurements. We have estimated photometric parallaxes for
these systems using the colour-magnitude relations defined in Paper I, combining 
individual estimates (including trigonometric data, as appropriate) using 
the precepts outlined in Paper VII. 

Two stars require particular comment. LP 320-16 (G 148-47) and G 148-48 form
a binary system, separation $\sim$4.5 arcseconds. Fleming (1998) obtained VI
photometry of this system as part of his follow-up observations of
candidate nearby M dwarfs from the ROSAT catalogue. He derived a photometric
distance of 11.2 parsecs for G 148-47, using the linear (M$_V$, (V-I)) calibration
derived by Stobie, Ishida \& Peacock (1989). However, no allowance was made
for the presence of G148-48, which was included in the photometric
aperture (Fleming, priv. comm. 2003).
We assume Fleming's measurements provide joint photometry
of the two stars. Based on inspection of the POSS II images and the 2MASS near-infrared
photometry\footnote{Note that the data listed in Table 2 for these two stars
are from the 2MASS All-Sky data release, which provides more accurate photometry
for stars in crowded environments.}, the two stars have very similar magnitudes and colours, and
we assume equal magnitudes in calculating the V magnitudes listed in Table 2. 
The revised distance estimate for the system is $\sim12.5$ parsecs.

Eighty-six stars from Table 2 are listed in the CNS3, including 51 that we
place within 20 parsecs of the Sun. In total,  
63 stars (in 62 systems) have formal distance estimates 
less than 20 parsecs, while 13 stars (from 12 systems) lie within 1$\sigma$ 
of the 20-parsec distance limit.

\section {New photometric observations}

Photometric follow-up observations of stars from both the NLTT2 and NLTT3 
samples were obtained between 2001 
December and 2002 July, using the St. Andrews photometer on the 1 metre 
telescope at the Sutherland station of the South African Astronomical 
Observatory. As discussed in more detail in Paper II and Paper VII, 
that photometer is equipped with a Hamamatsu R943-02 GaAs 
photomultiplier, and observations were made using a Johnson-Cousins BVRI filter set. 
Most measurements used a 21\arcsec\ diameter aperture, with a 
31\arcsec\ aperture employed for conditions of poor seeing. The relatively large 
aperture sizes lead to the inclusion of other stars in a few cases, as noted further below.

The observations were reduced using identical methods to those outlined in Papers II and VII
(see also Kilkenny \etal, 1998). The photometry was calibrated using 
both E-region standards (Cousins, 1973; Menzies \etal, 1989) and 
redder standards from Kilkenny \etall None of the program stars 
have previous observations, but we have demonstrated
that our photometry is consistent with the standard 
Cousins systems to better than 1\%. The formal uncertainties of our
program star measurements seldom exceed 2\%, and uncertainties in
the derived photometric parallaxes are dominated by the intrinsic width of the
main sequence ($\sigma = 0.25$ to 0.4 magnitudes - see Paper I).

Table 3 presents the results - observations of 77 stars from NLTT2
and 24 stars from NLTT3. In addition to 
the optical data, we list the positions and near-infrared photometry from 
the 2M2nd\footnote { Note that slight changes in both 
the astrometry and photometry of these objects may be present in the final 2MASS 
data release.}. Several stars with multiple measurements have unexpectedly high
residuals and may be variable at the 5-10\% level (see notes to Table 3). We
have estimated photometric parallaxes using the (V-I), (I-J) and (V-K)
calibrations given in Paper I, combining the individual estimates using the
same weighting scheme outlined in Papers II and VII.
Nineteen stars in Table 3 are formally within 20 parsecs, while
two others have photometric distances within 1$\sigma$ of that limit.

\section {Spectroscopy}

\subsection {Observations}

We have obtained moderate-resolution spectroscopic follow-up observations of
448 stars, including 120 stars from the NLTT1 sample, 
186 from NLTT2 and 142 from NLTT3. The observations were made by
several different observing teams using facilities at Kitt Peak National
Observatory (2.1- and 4-metre telescopes), Cerro Tololo Interamerican Observatory
(1.5- and 4-metre telescopes) and Apache Point Observatory (3.5-metre telescope).
All of the spectra cover the wavelength range 6,000-10,000\AA\ at a resolution 
of 2-3\AA. Most observations of NLTT2 and NLTT3 stars were made in
tandem with the NLTT1 observations described in Paper VII, and full
details of the instrumentation, observers and observing
conditions for those observing runs are given in that paper. However, the
present paper also includes observations from three more recent time
allocations, as follows:
\begin{itemize}
\item Reid and Cruz used the GoldCam spectrograph with the KPNO 2.1-metre telescope
on 2003 October 8, 9 and 11 to obtain spectra of 130 candidate nearby
stars, including most of the NLTT1 stars listed in this paper.
As before, the 400 line mm$^{-1}$ grating, blazed at 8000\AA, was employed 
with an OG550 order-blocking filter, giving spectra covering the wavelength
range 6000 to 10000\AA\ at a resolution of 4.3\AA.
Conditions were generally non-photometric, with high
cirrus often present. Observations were 
made using a 1\farcs5 slit, matching the seeing. 

\item Covey obtained spectra of eight stars using the Double Imaging Sectrograph (DIS II) on 
the 3.5-metre telescope at Apache Point Observatory on 31 October 2003. The 
medium resolution (300 line  mm$^{-1}$) grating was employed on the red camera, 
covering the 6000 to 10000\AA\ region at a
dispersion of 3.15 \AA pix$^{-1}$ and a resolution of 7.3\AA. 

\item Finally, Reid \& Cruz were able to observe 43 nearby star candidates
using MARS on the Mayall 4-metre telescope on February 11, 12 and 13 2004.
The VG0850-450 grism was used with a 1$\farcs$5 slit, giving 
$\sim6\AA$ spectral resolution. Conditions were clear and photometric for these observations.
\end{itemize}

All spectra were bias-subtracted, flat-fielded, corrected for bad pixels, 
extracted and wavelength- and flux-calibrated using the standard IRAF routines 
CCDPROC and DOSLIT. Wavelength calibrations were determined using He-Ne-Ar arcs 
taken at the start of each night. The spectra were flux calibrated using 
observations of standard stars from Oke \& Gunn, (1983) and Hamuy \etall (1994). 
At least one flux standard  
was observed each night, and comparison of repeated independently-calibrated 
observations of individual program stars indicates that the derived spectral 
energy distributions are consistent to better than 5\% over the full wavelength 
range. This is more than adequate for the purposes of this program.

As in previous studies, we have calculated a series of narrowband indices,
measuring the strengths of the more prominent molecular
absorption features. Indices measuring TiO (7020\AA\ band), 
CaH (6300 and 6800\AA), CaOH (6250 \AA), VO (7300 \AA) and H$\alpha$ 
are given in Table 4.
The bandpasses are defined in PMSU1 and by Kirkpatrick \etall (1999). 
As discussed in Paper VII, standard star measurements from our  
observing runs show good agreement with previously 
published data, with no evidence for systematic offsets and typical
r.m.s. residuals of $\sigma = \pm0.01$. Repeated observations indicate
that the program star measurements are generally accurate to $\pm0.02$.

Table 4 also lists spectral types. Those values are derived primarily from 
the TiO5 measurements (using the calibration given in paper III), although
we use the VO-a index and visual inspection to verify results for
the relatively small number of stars with spectral types later than M5. 

Positions and near-infrared photometry for the NLTT stars with
new spectroscopic observations are listed in Table 5. Four systems require
special mention:
\begin{description}
\item[G73-27/LP 469-118:] Luyten identified this as a close binary, separation 3
arcseconds. The system is unresolved in the 2M2nd dataset, but resolved in
the All-Sky data release (see Appendix I). The JHK$_S$ magnitudes listed in Table 5 are for 
2MASS 02081218+1508424 and 2MASS 02081238+1508443 from the latter source. 
The system was resolved at the telescope, and each component observed separately, 
and we estimate distances of 20.6 and 21.5 parsecs, respectively.
\item[LP 414-108:] This 2-arcsecond separation equal-magnitude binary was not
recognised as such by Luyten, who lists a spectral type of g-k for the system.
The photometry listed in Table 5 is for 2MASS 04103830+2002264 and 04103813+2002241
from the 2MASS All-Sky release.  
Spectroscopy of the individual components gives distance
estimates of 61.8 and 65.6 parsecs, respectively.
\item [G199-16:] The 2MASS image of this source appears slightly elongated, and
the 2M2nd data release deconvolves the data into two point sources: 
2MASSJ1229095+623938 (J=10.34, H=9.36, K$_S$=9.32) and 
2MASSJ1229096+622939 (10.16, 9.64, 9.37). In paper I, we noted that the 
former source has a (J-K$_S$) colour consistent with an M7 dwarf, although the
optical/IR colours suggested an earlier spectral type and the (J-H)/(H-K$_S$)
colours are anomalous. The 2MASS All-Sky release lists only a single source
at this position, with J=10.058, H=9.523, K$_S$=9.266 and colours consistent
with spectral type derived from our low-resolution spectroscopy. We did not
see any evidence for binarity at the telescope, so it seems likely that this
object is single. We estimate a distance of 29 parsecs.
\item [G 215-46/LP 188-2:] LP 188-2 is one of the photometric outliers listed in Appendix I. 
Luyten lists this star, m$_r$=16.9, (m$_r$-m$_{pg}$)=1.3, as a close companion of 
G215-46 (m$_r$=16.1, (m$_r$-m$_{pg}$)=1.3), with a separation of 1.5 arcseconds at $\theta=86^o$.
As noted in the appendix, there is no evidence for significant image elongation on
either the 2MASS scans or the POSS I/II photographic plates, and LP 188-2 may not exist. 
In any case, the trigonometric parallax of G215-46 (2MASSJ2226166+483730) indicates a 
distance of 32 parsecs (Table 1).  
\end{description}
With the exception of the last three systems, the 
near-infrared photometry listed in Table 5 is taken from the 2M2nd release.

\subsection {Metallicities, absolute magnitudes and distance estimates}

Narrowband indices provide a means of estimating both spectral types and
spectrophotometric parallaxes for the NLTT nearby-star candidates.
Paper III presents absolute magnitude (M$_J$) calibrations for several
indices, but those are valid only for stars with
near-solar metal abundance. Fortunately, we can verify the metallicities
of the NLTT dwarfs in the current sample by comparing the relative
strengths of the CaH and TiO indices against reference measurements of disk dwarfs and
of intermediate and extreme subdwarfs. Figures 3, 4 and 5 plot those data for
the current set of NLTT1, NLTT2 and NLTT3 stars, respectively. The reference
data for disk dwarfs are from the PMSU survey (PMSU1, PMSU2), while the
subdwarf measurements are from Gizis (1997). The overwhelming majority
of the program stars are clearly disk dwarfs, and we have have used the TiO5, CaH2 and CaOH indices to estimate absolute magnitudes, deriving
a weighted average as described in Paper VII. The resulting distance
estimates are listed in Table 5. 

Three stars from the NLTT2 sample lie well below the disk sequence in
the (CaH2, TiO5) plane: LP 16-197 (TiO5=0.34, CaH2=0.31); 
LP 466-156 (0.47, 0.39); and LP 881-272 (0.58, 0.46).  The TiO5
values correspond to disk dwarf spectral types M4.5, M3 and M2, respectively.
Considering the distribution in the (TiO5, CaH3) plane, LP 16-197 lies within
the main body of data and is likely to be a disk dwarf. In contrast, 
the other two stars are offset in CaH3, and 
we identify both as intermediate subdwarfs\footnote{Note that this classification indicates only that the stars are mildly metal-poor relatively 
to the local disk population (perhaps [M/H]$\sim-0.7$), and does not 
necessarily imply membership of the halo population.}. Their spectra are
compared to standard stars in Figure 6a.

Similarly, several stars lie below the disk sequence in the NLTT3 sample, but
only three stars LP 97-817 (TiO5=0.39, CaH2=0.33), G 174-25 (0.67, 0.49)
and LP 109-57 (0.59, 0.46) are offset in both TiO5/CaH2 and TiO5/CaH3. The
TiO5 measurements correspond to disk spectral types of M4, M1.5 
and M2, respectively, and their spectra are compared to standards in Figure 6b.
As with the NLTT2 stars, all three are likely to be intermediate subdwarfs.
We have assigned spectral types using the CaH2 calibration derived by Gizis (1997). 

All five intermediate subdwarfs are expected to be subluminous in M$_J$.
We can estimate the extent of this effect using 2MASS photometry of
late-type dwarfs and subdwarfs with known 
trigonometric parallaxes. Figure 7 plots the (M$_J$, TiO5) and (M$_J$, CaH2)
colour-magnitude diagrams - sdM and esdM stars fall below the disk main sequence
in these diagrams. Since the NLTT stars lie near the upper edge of the
sdM sequence in the TiO/CaH planes, we adopt offsets of $\Delta M_J$=+0.6
for LP 881-272, G174-25, LP 109-57 (sdM3) and $\Delta M_J$=+0.3 for LP 466-156
(sdM4) and LP 97-817 (sdM5). All five remain beyond the 20-parsec limit,
although LP 97-817 has a formal distance estimate of only $\sim24$ parsecs.

Fifty-seven stars, all from NLTT2, have both VRI photometry (Table 2) and
optical spectra. Figure 8 compares the photometric and spectrophotometric
distance estimates. The agreement is within the formal
uncertainties, with a mean offset of -0.11 magnitudes 
(photometric-spectrophotometric) and a dispersion of 0.36 magnitudes. 

\section {A local census: K and early-type M dwarfs}

The main goal of the present program is to compile a census of late-type
dwarfs within 20 parsecs of the Sun. 
With the addition of the observations described in this paper, we have 
Essentially completed our NLTT follow-up observations. We have
trigonometric, photometric or spectroscopic parallax estimates for 
all save three of the 1913 nearby-star candidates in the NLTT1, NLTT2 and NLTT3
samples\footnote{ The three stars which still require observation are: F I-325,
NLTT\#23767, m$_r$=11.1, (m$_r$-K$_S$)=3.6; Grw +74:5033, NLTT\#37764, 
m$_r$=11.8, (m$_r$-K$_S$)=4.4; Grw+69:3815, NLTT\#21156, m$_r$=11.9, 
(m$_r$-K$_S$)=5.0.}. Eight hundred and fifteen stars are placed
within 20 parsecs of the Sun, including 312 additions to 
nearby-star catalogues. A further 224 stars are estimated to lie
within 1$\sigma$ of the 20-pc boundary of our NStars survey. 

The three NLTT samples discussed so far in this series of papers
are limited to colours (m$_r$-K$_S)>3.5$ magnitudes. This
corresponds to stars with M$_V >11$, or spectral types later than
$\approx$M3. This section explores ways of extending coverage 
through spectral type K, using the NLTT together with previously-published 
nearby-star catalogues. This also allows us to identify
nearby M dwarfs, particularly stars with low proper motions,
which have escaped inclusion in the NLTT samples.

\subsection {Nearby early-type M dwarfs in the NLTT}

The optical/infrared selection criteria used to isolate 
candidate nearby stars in the NLTT were aimed primarily at mid- and
late-type M dwarfs. However, in Paper VII, we matched those criteria 
against data for known M dwarfs catalogued in the PMSU survey, including
spectral types M0-M2.5. That comparison shows that the linear relation
adopted at the bluest colours, (m$_r$-K$_S) < 5$, provides an effective
means of selecting stars within 20 parsecs with (m$_r$-K$_S)>$2.0 magnitudes,
1.5 magnitudes bluer than the cutoff adopted in constructing the NLTT1, NLTT2
and NLTT3 samples. We have therefore compiled a list of K and early-M 
nearby-star NLTT candidates by selecting stars from the rNLTT with colours in 
the range $2.0 < (m_r-K_S) \le 3.5$ and apparent magnitudes that
meet the criterion, $m_r < 2.17(m_r-K_S)+3.65$ (Paper I). 

Four hundred and two NLTT stars fulfil these criteria (Figure 9, upper panel).
Three hundred and eighty-eight of these candidates 
are either included directly in the Hipparcos catalogue or are companions of 
Hipparcos catalogue stars, and most have accurate 
trigonometric parallax measurements. Only 71 stars in 69 systems (all
save seven listed in the CNS3) have
formal distance estimates of less than 20 parsecs (Table 6).  
Four binary companions of these stars (Gl 282B, Gl 414B, Gl 420B and Gl 425B)
are already included in the NLTT1/2/3 samples; 
a further 15 unresolved companions also need to be taken into account.
In two cases, BD+14:251 and Gl 225.2ABC, the Hipparcos parallax has
a substantial uncertainty ($>30\%$), reflecting the influence of
close optical or binary companions. As discussed further below, 
photometric parallax measurements suggest that both systems lie well 
beyond 20 parsecs.

Table 7 lists relevant data for the 14 stars that lack Hipparcos data.
Most have no direct trigonometric
parallax data, but the available distance estimates, pricipally the (V-K$_S$)
photometric parallax, place all save four well beyond 20 parsecs. 
The remaining stars include the eclipsing binary system DV Psc (but see further below), both components of HD 95174, and G204-45. None of these systems 
is listed in the CNS3. 

The lower panel of figure 9 plots the (M$_V$, (V-J)) colour-magnitude
distribution for stars with $\pi_H > 50$ mas (Table 7) and stars with no
Hipparcos data (Table 6). We include data for 14 fainter companions of 
those stars. Several stars fall well below the main sequence. 
BD+14:251, Gl 225.2B, Gl 225.2AC and DV Psc all lack accurate trigonometric 
parallax measurements. There is no evidence that any of these stars is
genuinely sub-luminous, and these three systems probably have
distances of 30-40 parsecs.  We have excluded them from the 20-parsec sample. 
In the case of GJ 1181A, the trigonometric parallax appears secure, and the
problem probably rests with current estimates of the relative photometry of the 
two components. 

Retaining GJ 1181, this sample of K and early-type M NLTT dwarfs
adds 80 stars in 69 systems to the 20-parsec census.
Figure 9b shows that the (m$_r-K_S$)=2.0 colour limit adopted 
for this sample 
corresponds to M$_V \sim 6.5$, or spectral type $\approx$K1. These
stars therefore bridge the gap between the later-type dwarfs in 
the NLTT1/2/3 samples and the more luminous samples discussed further
below.

\subsection {Additional nearby stars}

We have searched the literature for cool dwarfs that fall 
in the area covered by the 2M2nd release and are within 20 parsecs, 
but are not included in our NLTT sample. The additions are drawn from
three main sources: the PMSU Survey; 
the Hipparcos catalogue; and the recent proper motion surveys by
L\'epine and collaborators. We concentrate on K and M dwarfs, and 
postpone discussion of ultracool dwarfs,
including results from our own survey, to future papers.

\subsubsection {The PMSU survey}

Most of the additions to the 20-parsec census are drawn from the Palomar-Michigan State
University M-dwarf spectroscopic survey (PMSU1, PMSU2). 
That survey used narrowband indices, derived from moderate-resolution 
optical spectra, to compute spectroscopic parallaxes
for $\sim2000$ M dwarfs catalogued in the CNS3, supplementing available
astrometric and photometric data. As discussed in Paper VII, 468 of the 
502 PMSU stars that are both in the NLTT catalogue, and have distance
estimates $d_{PMU} < 20$ parsecs, meet our colour-magnitude
selection criteria and are included in the NLTT1/2/3 samples. However, 
there are a further 137 M dwarfs in the PMSU catalogue which also have 
formal distance estimates less than 20 parsecs, but are not included in 
the present sample for a variety of reasons. 

Table 8 lists relevant data for the additional PMSU stars. 
Those stars fall into five categories. First, as discussed in Paper VII, 
27 stars
are not in the NLTT catalogue, and therefore are excluded {\sl a priori} from
our sample. Second, 39 NLTT dwarfs with formal distance
estimates of less than 20 parsecs fail our colour/magnitude selection criteria
(m$_r$ too faint, or(m$_r$-K$_S$) colours too blue).
Thirty-four of those stars were included in the Paper VII statistical analysis.
Three of the five additional stars (GJ 1054B, LHS 2658 and G273-186A) have
d$_{PMSU} > 20$ parsecs, but have companions with d$_{PMSU} < 20$. The 
remaining two stars are HD 129715, a K7 dwarf which was not included 
in PMSU but has a spectroscopic parallax of 54 milliarcseconds (mas), and 
its M4.5 companion, LP 858-23,
which has $d_{PMSU}=22\pm4$ parsecs. Given the proximity of
the latter star to the M3/M4 break in the colour-magnitude diagram, we adopt 
54 mas as the parallax estimate for both stars. Note that several other stars in
the `too blue' NLTT category also lie near this pronounced discontinuity in the
main sequence. 

The third category includes 34 brighter dwarfs, eliminated
from the NLTT sample due to
saturation in one or more passband (usually H) in the 2M2nd release. 
The 2MASS All-Sky release includes analysis of shorter exposure scans, 
allowing photometry of bright stars, and we list final release 
data for those stars in Table 8. 
The earlier-type stars in this dataset (such as Castor and Pollux) 
are included because they are companions of later-type dwarfs in the
20-parsec census. 

Half a dozen PMSU stars listed in the NLTT were omitted from our sample because
neither our NLTT cross-referencing nor the rNLTT succeeded in identifying the
correct 2MASS counterpart. 
Finally, 39 PMSU stars are secondary/tertiary companions,
merged with the primary star in the 2MASS scans. Some of these
systems have high resolution images at near-infrared
wavelengths, but in most cases we have used the relative optical
magnitudes, mainly from the CNS3, with standard colour-magnitude
relations to estimate the relative flux at near-infrared wavelengths,
and deconvolve photometry of the individual components from the 2MASS data. 

\subsubsection {The Hipparcos catalogue}

The Hipparcos satellite provided milliarcsecond-accuracy astrometry for
110,000 pre-selected, bright (V$<13$ mag.) stars, including most
stars listed in the CNS3. 
The resulting catalogue (ESA, 1997) is statistically complete to 25
parsecs for stellar systems with components brighter than 
M$_V = 9$ (Jahreiss \& Wielen, 1997). We can use this dataset to 
provide complete coverage of K dwarfs in the 2M2nd survey. 
We have identified all Hipparcos stars with M$_V > 5.0$ and 
parallaxes $\pi_H > 50$ mas, and cross-referenced that sample against 
the 2M2nd release. Most late-K and M dwarfs are already in the 
20-parsec sample. The majority of the additions are included in the 
CNS3 (but not PMSU), and are sufficiently bright that the 2M2nd near-infrared
magnitudes are saturated. As with the bright PMSU-selected stars, we 
have used the 2MASS all-sky
release to obtain JHK$_S$ for these stars. A number of stars have binary
companions, and we have added those to the census, deconvolving infrared
magnitudes for the individual components where necessary.

Relevant data for the Hipparcos-selected 20-parsec stars are listed in 
Table 8. Eleven of the 62 stars listed there 
were identified directly by Hipparcos as members of the  
Solar Neighbourhood. However, not all stars with $\pi_H > 50$ 
mas are confirmed as immediate neighbours. We have excluded two
such stars from the present sample: HIP 114110 (BD-15:6346B) and HIP 114176
(CD -32:17393B). In both cases, the formal uncertainty in the measured parallax is
$\sim10\%$. Each is a secondary star in a binary system: BD -15:6346, or HD 218251, 
has an Hipparcos parallax of $\pi_H=13.26\pm1.01$ mas; CD -32:17393, or HD 218318, has
$\pi_H=13.01\pm1.00$. It is likely that the 
apparently large parallaxes measured for the two secondary stars are spurious,
perhaps due to the proximity of the bright primary. 

\subsubsection {The LSR surveys}

The proper motion surveys undertaken by L\'epine, Shara \& Rich (2002a, 
2002b, 2003 - hereinafter LSR1, LSR2 \& LSR3) are a
modern reprise of Luyten's all-sky surveys. Rather than use visual 
or automated plate-blinking techniques, 
proper motion stars are identified through digital
subtraction of scans of first- and second-epoch Palomar and UK/AAO Schmidt
sky survey plates. To date, the published catalogues are limited to
northern hemisphere stars with motions exceeding 0.5 \pma. 
We cross-referenced their low galactic-latitude sample (LSR1) against
the 2M2nd database in Paper VI (Reid, 2003). Most of the newly-identified
proper motion stars are either white dwarfs or halo subdwarfs. We have
estimated distances for the disk dwarfs in the sample using either the
spectral-type/M$_J$ relation (Paper V) or, if spectral types are not
available, the (R-K$_S$) colours. Similarly, we have cross-referenced the
high-latitude discoveries from LSR2 against the 2M2nd, and estimated
distances using the same techniques. Table 9 lists LSR stars likely
to lie within 20 parsecs of the Sun. More detailed information is available for
two stars: LSR1826+30, a high-motion (2.38 \pma) star, has
broadband VRI photometry (LSR3); and LSR1835+32 is 2M1835+32, which has
a measured trigonometric parallax (Paper IV). 

\subsection {The 20-parsec census}

Combining the early-type dwarfs listed in Table 6 and the literature data
from Table 8 with our NLTT sample gives a total of 1027 stars,
in 890 systems, within 20 parsecs of the Sun. Figure 10 plots the
absolute magnitude distribution of the full dataset, identifying the 
contribution from stars previously catalogued in the CNS3. Since we are
considering stars on an individual basis in this figure, we have not 
applied any statistical corrections to the absolute magnitudes 
(see below, \S7.3).

As discussed in Paper VII, the main impact of the present survey lies at
fainter magnitude, M$_J > 7$, although we note that our study
also provides improved, self-consistent distance estimates for a number of 
brighter stars. The dataset discussed in this paper adds relatively few 
stars at the faintest magnitudes, M$_J > 10.5$ (spectral types later 
than M8), but that is not unexpected given the limiting magnitude of 
the NLTT survey. 

\section {The Luminosity Function}

One of the main goals of this project is a
statistically-improved determination of the luminosity function for
low-mass stars and brown dwarfs. With the effective completion of our 
NLTT follow-up observations, and the addition of complementary data 
discussed in the previous section, we can construct a preliminary
luminosity function, and examine the likely completeness of the
current census. The following sections summarise stellar properties
at near-infrared wavelengths, and identify two reference samples 

\subsection {The main sequence in M$_J$}

Most previous luminosity function studies have centred
on the distribution at visual absolute magnitudes, $\Phi$(M$_V)$. We lack 
accurate V-band photometry for many stars in the NLTT
sample, precluding direct derivation of that quantity. 
On the other hand, we have, by construction, 2MASS JHK$_S$ photometry for 
all stars in our 20-parsec census, and the photometric and
spectrophotometric luminosity relations applied to late-type
dwarfs are calibrated against M$_J$. Under these circumstances, it 
clearly makes sense to construct a near-infrared luminosity function.

Figure 11 provides an overview of stellar properties in this less
familiar r\'egime. We plot the (M$_J$, (V-J)) and (M$_J$, spectral type) 
diagrams for a sample of single stars with accurate photometry and 
trigonometric parallaxes measured to an accuracy better than 5\%. 
The main sequence shows significant dispersion ($sigma \sim 0.35$ mag)
brighter than the M3/M4 discontinuity, which occurs at M$_J\sim8.5$ (see papers
I \& III for further discussion of this feature). As in all other planes, the
main sequence narrows significantly at later spectral types. 
Table 6 provides a coarse reference guide to average colours and
spectral types as a function of M$_J$. 

\subsection {Calibrating J-band statistics: the 8-parsec sample}

The northern 8-parsec sample includes stars and brown dwarfs with 
distances d$<$8 pc and declinations $\delta \ge -30^o$. This dataset
provides a baseline guide to the space densities of late-type
dwarfs in the Solar Neighbourhood, and hence the expected numbers of 
stars in the 20-parsec census. The sample was defined originally
by Reid \& Gizis (1997), and subsequent modifications 
are summarised by Reid \etall (1999) and in Paper IV of this series. 

The present analysis takes two
further corrections into account: Gl 831 AB is eliminated, based on 
the parallax of 0.1175 arcseconds derived by S\'egransan \etall (2001) in 
their re-analysis
of Hipparcos data, allowing for orbital motion; and Teegarden \etall (2003) have
discovered an M6.5 high proper-motion star with a likely distance of
$\sim3.6$ parsecs. We note that Ducourant \etal (1998) have published a 
trigonometric parallax measurement of $156\pm4$ mas for 
the M5 dwarf, G 180-60, 
formally placing that star within our sample. However, at that distance,
G 180-60 falls $\approx$1 magnitude below the main-sequence\footnote{ Similar situations prevail for the other nearby-star candidates 
(P1, P5 and P18) identified in the Ducourant \etall paper.}. Since the
spectrum shows no evidence for low metal abundance (PMSU1), we regard 
the previous photometric and spectroscopic distance estimates (10-12 parsecs) as more reliable. 

Seven stars in our survey have formal distance estimates of less than 8 parsecs:
\begin{description}
\item [LP 467-16:] An M5 dwarf, catalogued in the CNS3 with a photometric 
parallax of $118\pm21$ mas and a PMSU spectroscopic distance estimate of
$10.5\pm3$ pc. Our revised estimate, based on multicolour photometry (Paper I),
is $7.6\pm1.2$ pc, corresponding to M$_J=9.68$.
\item [LP 993-116:] Lying at -43$^o$, this star is well south of the -30$^o$
declination limit of the 8-parsec sample. We estimate a photometric distance of 
$7.0\pm1.7$ parsecs (Paper II) and M$_J$=8.85. However, LP 993-116 is a wide 
companion of BD-44:836, for which we estimate a photometric
distance of $10.6\pm1.6$ parsecs (Paper II). With (V-I)=2.99, LP 993-116 
lies near the M3/M4 break in the main sequence, and the longer distance estimate 
to the earlier-type BD-44:836 ((V-I)=2.72) is likely to be more reliable.
\item [LHS 6167:] Also known as G 161-7, this mid-type M dwarf is
not listed in the CNS3. We derive a photometric distance of $6.7\pm1.3$ pc
(Paper II), corresponding to M$_J=9.47$. 
\item [G 161-71:] Another new candidate from our SAAO observations (Paper II),
 we derive a distance estimate of $6.2\pm1.2$ pc and M$_J=9.51$.
\item [G 165-8:] Listed in the CNS3 with a parallax of $126\pm22$ mas, the 
PMSU distance estimate is $10.5\pm3$ parsecs. Combining the available
trigonometric, photometric and spectroscopic data (this paper), we derive
 $7.9\pm0.8$ parsecs and M$_J$=8.12 for this M4 dwarf.
\item [LP 71-82:] An M4.5 dwarf, our spectroscopic parallax gives a distance
estimate of $6.9\pm1.4$ parsecs and M$_J=9.35$ (Paper VII). 
\item [LP 876-10:] Our SAAO BVRI photometry leads to a photometric
distance of $7.2\pm1.4$ parsecs and M$_J=8.81$ (Paper VII). 
\end{description}
All of these stars are mid-type M dwarfs (spectral types M3 to M5), lying near
the break in the main sequence relation, and accurate trigonometric parallax
data are required to verify the photometric/spectroscopic distance estimates.

Only three of the seven new 8-parsec candidates 
(LHS 6167, G 161-71 and LP 71-82) are retained in the luminosity function
analysis; the Malmquist-corrected distances (see below, \S7.4) for the 
remaining northern stars all exceed 8 parsecs. Thus, integrating all 
changes since the original definition of the northern 8-parsec sample 
gives a net result of stasis: 13 stars in 10 systems have been eliminated 
from the original sample, while 13 new stars (in 10 systems) have been added. 

The current northern 8-parsec sample therefore consists of 140 main-sequence
stars (including the Sun), three brown dwarfs (Gl 229B, Gl 570D and LP 944-20)
and nine white dwarfs in 108 systems. 
The overall multiplicity, ${{N_{bin} + N_{triple} +..} \over N_{sys}}$, is 29\%. 
Only four systems lack accurate (${\sigma_\pi \over \pi} < 7\%$)
trigonometric parallax data; all of the resolved main-sequence stars have 
direct JHK$_S$ photometry, while 
there is sufficient ancillary data to estimate M$_J$ for unresolved close
binaries. Figure 12 plots the visual and J-band luminosity
functions derived from this data. We identify the separate contribution 
from primaries/single stars and secondary components, and the error bars
show the formal Poisson uncertainties derived from the total counts in each bin. 

\subsection {Earlier spectral types: The PMSU4 25-parsec sample}

With a total of only 152 objects, the 8-parsec sample provides 
sparse statistics for stars more massive than the Sun. 
While the present project centres on K and M dwarfs, data for earlier-type
stars are useful in setting those results in context. 
Reid, Gizis \& Hawley (PMSU4) have compiled an Hipparcos 
all-sky 25-parsec census to M$_V=8.0$ (M$_J < 4.5$), adding 
photometric data for lower-luminosity, main-sequence companions that
lack separate Hipparcos entries. The PMSU4 sample
is drawn from a volume approximately four times that of the present 2M2nd census
(all-sky to 25 parsecs versus 48\% of the sky to 20 parsecs), and forty times
that of the northern 8-parsec sample. As a consequence, these data provide
a more reliable reference for stellar space densities at bright absolute magnitudes.

The PMSU4 sample includes 1024 main-sequence stars in 764 systems.
The overall multiplicity fraction is only 30\%, significantly lower than 
the value of 44\% derived for solar-type stars by Duquennoy \& Mayor 
(1991), suggesting that as many
as 150 binary companions remain to be discovered. However, 
the overwhelming majority of those stars are likely to be fainter than M$_V=8$.

Most of the PMSU4 Hipparcos stars have direct JHK$_S$ photometry from the 
the all-sky 2MASS data release. We have used optical photometry
(usually B, V) to estimate M$_J$ for the brightest stars, saturated
in even the final 2MASS catalogue, and for unresolved companions (see Reid 
\& Gizis, 1997 for further details). Individual stars have
absolute magnitudes spanning the range  $-1 < M_J < 8.5$, but the sample
clearly becomes incomplete at M$_J > 5.5$. The PMSU4 dataset therefore 
provides a semi-independent\footnote {Hipparcos stars lying within the
area covered by the 2M2nd and with parallaxes $\pi_H > 50$ mas are common
to both samples.} check of the completeness of the 20-parsec census at
those magnitudes. 

\subsection {The J-band luminosity function, $\Phi$(M$_J$)}

As described in the previous section, our 20-parsec census includes 1027 stars
in 890 stellar systems. Before deriving the luminosity function from this
dataset, we need to make due allowance for systematic statistical 
bias present in the absolute magnitudes estimated for individual stars. 
Those corrections are applied on a system-by-system basis. In the case of
systems with trigonometric parallax data, the corrections are small, 
since the astrometric accuracies are typically 2\% or better. This corresponds
to Lutz-Kelker bias of less than 0.05 mag. in M$_J$ (Lutz \& Kelker, 1973).

However, while 544 systems have trigonometric 
parallax measurements, the distance estimates for over 340 NLTT/PMSU stars are
derived from either photometric or spectrophotometric parallaxes. The
latter stars are effectively drawn from a magnitude-limited 
sample at each spectral type or colour, so it is appropriate to use 
Malmquist bias (Malmquist, 1936) to adjust the individual absolute 
magnitudes for statistical analysis. Defining M$_0$ as the true absolute
magnitude and M$_{obs}$ as the observed absolute magnitude for a given
star, we have
\begin{displaymath}
M_0 = M_{obs} - { {\sigma^2 \over {log_{10}e}} {{d log A_S} \over {dm}} }
\end{displaymath}
where $\sigma$ is the uncertainty in M$_J$, and ${{d log A_S} \over {dm}}$
the slope of the logarithmic number-magnitude distribution for a given
colour/spectral type. The last parameter varies significantly with
apparent magnitude for a proper-motion limited sample, but is close
to 0.6/magnitude (a uniform-density distribution) for the brightest/nearest
stars sampled in our survey. As noted above,
we set $\sigma$ to match the dispersion in the calibrating
colour/magnitude relations: 0.25 magnitudes for low-luminosity stars 
((I-J)$>$ 1.75); 0.4 magnitudes for stars above the M3/M4 break in the 
main-sequence ((I-J)$<$1.4); and 0.5 magnitudes at the break. Thus, for
\begin{displaymath}
M_0 = M_{obs} - 1.38 \sigma^2
\end{displaymath}
the absolute magnitudes are corrected by -0.09, -0.22 and -0.35 magnitudes,
respectively.

Figure 13 compares the J-band luminosity function derived from our
20-parsec census against corresponding results for the northern 
8-parsec sample and the PMSU4 25-parsec sample. Since the present
survey covers 48\% of the sky, while the 8-parsec survey covers 75\% of the
sky, there is a factor of 10 difference in the sampling volume. 
Since this is a statistical comparison, Malmquist and Lutz-Kelker bias are 
taken into account, and we show the relative contribution of systems (single
stars and primaries) and companions to both datasets.

Clearly, despite the new discoveries from our NStars project,
the current 20-parsec census is far from complete. 
Based on the 8-parsec sample, we expect $\sim1290\pm110$ stars in 
$\sim920\pm96$ systems over the absolute magnitude range $4 < M_J < 11$,
where the uncertainties reflect the Poisson statistics of the
local sample\footnote{ As a reference, the predicted numbers in this
magnitude range for an all-sky 20-parsec survey are 
2690 stars in 1915 systems. Adding earlier spectral types, we 
expect 2915 main sequence stars in 2125 systems.}. In contrast, 
the current 20-parsec census includes only 853 stars 
($\sim66\%$ completeness) in 727 systems ($\sim 80\%$ completeness).

Comparing the predicted and observed results in more detail, several 
points can be made:
\begin{description}
\item[late-G/K dwarfs, $4 < M_J < 6$:] there is good agreement between 
the space densities derived from the PMSU4 and 20-parsec samples 
at these magnitudes. The current 20-parsec census becomes incomplete 
at M$_J$=4.25, reflecting the M$_V>5$ selection imposed on the
Hipparcos sample, while the PMSU4 dataset becomes incomplete at M$_J > 5.5$ (M$_V > 8$). Note that the 8-parsec number densities at $4 < M_J < 5$
are higher that the corresponding densities derived from the
PMSU4 sample. The former values, however, are based on only 7 and 6 stars  
at M$_J$=4.25 and 4.75, respectively, so the differences are equivalent
to offsets of $\sim1.5\sigma$.

\item[M dwarfs, $11 > M_J >6$:] most of the deficit between the observed 
and predicted number counts resides at faint magnitude, M$_J > 7$ (indeed, the
20-parsec sample shows a $\sim1\sigma$ excess at $6 < M_J < 7$). Based on
the 8-parsec data, we predict 910 stars in 630 systems at magnitudes
$7 < M_J < 10$; this compares with 547 stars in 461 systems in the present
census. This suggests that, at these absolute magnitudes, approximately 25\% of 
the stellar systems and 40\% of the individual stars are missing from our 
census. The most substantial shortfalls in density ($>2\sigma$) lie 
at intermediate magnitudes, $7 < M_J < 8.5$ ($10 < M_V < 13$, spectral types M2 to M4). 

\item[Binarity:] the binary fraction ($N_b \over N_{sys}$) in our 
20-parsec census is only 15\%, significantly lower than balue measured for
the 8-parsec sample (35\%). 

\item[Proper motions:] only $\sim$4\% of the stars in the current 20-parsec 
sample have motions below the NLTT limit, $\mu < 0.18$ \pma. This
is a factor of two to three lower than expectation, based on the measured
velocity dispersion of local disk dwarfs (PMSU4).

\end{description}

The main conclusions that we draw from this comparison are, first,
that significant numbers ($\sim150$) of low-luminosity companions 
remain to be discovered within the 20 parsec sample. 
This result, echoing the PMSU4 study, is not surprising,
since at least half of the binaries in the 8-parsec sample would be unresolved
at distances approaching 20 parsecs. Second, between 150 and 200 stellar
systems, mainly mid-type M dwarfs, are missing from the current census. 
Third, the relatively
good agreement between the 8-parsec and 20-parsec number densities for
systems with $8.5 < M_J < 10.5$  suggests that relatively few 
($\sim20$?) low-luminosity 
M-dwarf {\sl systems} remain to be discovered within the regions covered by
the present survey\footnote {We note that our
ultracool dwarf survey (Paper V) fills in the apparent deficit 
between the 20-parsec and 8-parsec space densities at 
M$_J > 10.5$ in Figure 12.}. 
Finally, it is likely that at least 60-70 stars with proper motions
$\mu < 0.18$ \pma are missing from the current census.

\section {Completing the census: a comparison of search methods}

Traditionally, nearby star catalogues have been constructed 
through a four step process. Wide-field proper motion surveys, coupled 
with rudimentary photometric data, provide the first cut, flagging 
thousands of systems with substantial tangential motions. 
Photometric and spectrophotometric follow-up observations are used to select 
the most likely candidates. Trigonometric parallax surveys, such as the USNO
program (Monet et al, 1992; Dahn et al, 2002) and the CTIOPI survey (Henry
et al, 2003), generally target the latter stars, providing more accurate 
distance measurements. With the exception of surveys like Hipparcos,
parallax programs rarely discover new nearby systems. Finally, 
high-sensitivity, high-resolution imaging and spectroscopy of those same targets
are required to complete the local census by searching for lower-mass/luminosity companions. 

Our survey fits squarely in this mold.
Starting with 23,795 NLTT stars overlapping the 2M2nd survey, we have 
selected 1913 sources with colours consistent with nearby main-sequence 
M dwarfs, and used follow-up BVRI photometry and spectroscopy to trim
that sample to 815 stars likely to lie within 20 parsecs of the Sun.
As noted in the introduction, we expect some incompleteness in the
final sample, since 10-12\% of the local disk stars have tangential motions
lower than the proper motion cutoff of the NLTT. We have 
identified approximately
one third of the low-motion stars from other sources, but even so, our
present census is incomplete by 20\%. 

What are most effective methods of finding the missing stars?
To answer that question, we need to consider the likely characteristics of
those objects. Based on the luminosity function comparison in the previous
section, we can divide them into three broad groups: $\sim$150 to 180 
companions of known (or as-yet undiscovered) nearby stars; 20-30 cool,
isolated dwarfs (spectral types M4.5 to M8); and $\sim150$ mid-type M
dwarfs (spectral types M2 to M4). Each requires a somewhat different approach.

In the case of the missing companions, both the targets and the search 
techniques are obvious. These objects can be discovered through high resolution
imaging and high-accuracy radial velocity monitoring of 20-parsec stars that
currently lack such data. These are straightforward, if observationally
intense, programs. 

Large-scale surveys, such as the NLTT, are a prerequisite for identifying 
isolated systems. The NLTT itself is known to be incomplete in two
important respects at faint magnitudes: the primary source catalogue at
southern declinations ($\delta < -40^o$) is the Bruce proper motion
catalogue, limited to m$_{pg} \sim 15$; and the high star density near the
Galactic Plane complicated the identification of fainter proper motion
stars on the Palomar plates. 
Both effects are clearly evident the ($\alpha, \delta$)
distributions plotted in Figure 1 of Paper I, and both percolate, to
some extent, to the present 20-parsec census, although the bright limiting
magnitude in the deep south is less important, since only 3\% of the
2M2nd lies south of  $\delta = -40^o$.

Figure 14 shows the latitude distribution of
the 20-parsec stars, divided into three absolute magnitude intervals: $M_J < 7$; 
$7 \le M_J < 9$; and $M_J \ge 9$. Since the 2M2nd has a complex distribution
on the celestial sphere, we have binned the stars in 10-degree intervals
in latitude, and calculated the fraction in each zone. As a reference, we 
plot the same distribution for NLTT stars with $8 < m_r < 10$ and
2M2nd counterparts - those stars are sufficiently bright that they do not
suffer from either selection effect. There is some evidence for a deficit
near the Plane at the faintest luminosities in the 20-parsec sample,
although the shortfall corresponds to only $\sim10$ systems.

New proper motion surveys, such as those by L\'epine \etal (2003), can 
avoid these systematics to some degree, since better plate material and
more sophisticated search techniques, such as digital image subtraction,
are now available. However, crowding can still lead to problems in 
finding faint proper motion stars in the highest star-density regions.
Moreover, just as with the NLTT survey, supplementary photometric 
(or spectroscopic) data are required to unequivocally identify Solar Neighbourhood members.

In principle, optical/infrared photometric surveys offer the best chance of
finding the later-type dwarfs - later than M4, the main sequence has low
dispersion, with colour increasing monotonically with decreasing absolute
magnitude, as illustrated in Figure 10. In practice, the effectiveness of
this approach is limited by the availability of accurate optical photometry -
particularly at low Galactic latitudes and in the south. SDSS, for example,
covers only 1 steradian, at high latitudes and predominantly in the northern
hemisphere. GSC2.2, derived from the digitized Palomar and UKST sky surveys,
offers a potential alternative, but that survey offers photometry of
only moderate accuracy ($\sim0.1$ to 0.15 mag.), and the current 
catalogue includes multiple (uncorrelated) measurements of stars 
with significant proper motions (from plates taken at different epochs). 
The 2MASS database lists optical photometry for sources within 5 arcseconds
of the infrared source, but those data (from the USNOA catalogue) are derived
from first-epoch sky survey material, and are therefore absent for sources
with moderate proper motion. Clearly, the 2MASS near-infrared photometry
(notably (J-K$_S$)) offers an alternative search method, which we have 
exploited in the other main thread of the current project (see Paper V
and Paper IX). That search technique, however, is effective only for dwarfs 
with spectral types later than M8\footnote {M7 dwarfs may prove particularly
tricky to locate, since they do not have distinctive near-infrared colours,
but are sufficiently faint at optical wavelengths to have eluded previous
surveys, like the NLTT.}.

Surveys for the missing mid-type M dwarfs are further complicated by the
fact that those stars lie near a break in the main-sequence. The 
substantial width (in absolute magnitude) at these spectral types (see
Figure 10) leads to substantial Malmquist bias, and a correspondingly 
high proportion of distant interlopers in any photometrically-selected
sample. 

Under these circumstances, a direct photometrically-based search for the
missing 20-parsec stars is not likely to be fully effective. We are therefore
pursuing a hybrid approach, taking advantage of the optical data present
in the 2MASS catalogue. First, we can use the USNO photometry to
search directly for 2MASS sources with optical/infrared colours consistent
with nearby M dwarfs; these should include stars with low tangential
motions, ($\mu < 0.15$\pma). Second, we can couple the {\sl absence} of an
optical counterpart in 2MASS with appropriate near-infrared photometric
criteria, and identify nearby M dwarfs with significant proper motions. As
indicated in \S1, we are currently pursuing both techniques, and will discuss
the results in a future paper in this series.

Finally, we should emphasise that the current 20-parsec census is not
immutable. It is likely that some of the stars added through the present
program are unresolved binaries, with correspondingly overestimated
photometric or spectroscopic parallaxes. We need to refine the current distance
estimates and, in particular, acquire accurate trigonometric parallax
measurements for the 40\% of the sample that currently lack such data. 

\section {Summary and conclusions}

As part of our continuing survey of the late-type dwarfs in the immediate
Solar Neighbourhood, we have presented photometric and spectroscopic
observations of over 800 proper motion stars from the NLTT catalogue.
These stars are drawn from three separate compilations, matching the
NLTT against 2MASS point-source data from the Second Incremental Release.
Two of those compilations are own work, matching sources with positional
coindence less than 10 arcseconds (NLTT1) and 60 arcseconds (NLTT2); the
third (NLTT3) is derived from the independent analysis by Salim \& Gould (2003).
In each case, the nearby star candidates are selected using location in
The (m$_r$, (m$_r$-K$_S$)) plane. There are a total of 1237, 369 and 307
nearby-star candidates, respectively, in the final three samples -
a total of 1913 stars. 
 With the addition of the observations presented in this paper, we
have compiled astrometric, photometric and/or spectroscopic distance 
estimates for 1910 stars. We identify 815 of those stars
as likely to be within 20 parsecs of the Sun. 

The nearby stars identified from our NLTT follow-up observations are
M dwarfs, predominantly spanning spectral types between M2 and M7.
We have extended coverage to earlier spectral types (late-G, K and early-M),
coupling colour-selection from the NLTT with literature data, notably stars
from the Hipparcos catalogue. The resulting 20-parsec census, drawn from
the 48\% of the celestial sphere covered by the 2M2nd, spans the absolute\
magnitude range $4 < M_J < 10.5$ and includes 1027 stars in 890 systems.

We have computed the J-band luminosity function, $\Phi_(M_J)$, of our
20-parsec sample, and compared the results against reference data for 
the northern 8-parsec sample, spanning the full stellar mass range (see
Paper IV), and an Hipparcos-based analysis of AFGK stars within 25 parsecs
of the Sun (PMSU4). In passing, we note that the present survey produced
only three additions to the former sample. The comparison indicates that
the 20-parsec census is complete to M$_J\sim 7$. At fainter magnitudes, 
and relative to the 8-parsec predictions,
there is an apparent deficit of $\sim170$ stellar systems ($\sim25\%$\
incompleteness) and $\sim360$ individual stars ($\sim40\%$ incompleteness)
at $7 < M_J < 10.5$ . 
At least half of the missing stars are companions of
known nearby stars, while approximately 90\% of the missing systems 
are likely to be concentrated between M$_J=7$ and M$_J=8.5$ - 
spectral types M2 to M4. 

We have discussed a number of techniques that could be used
to supplement our current survey and complete the 2M2nd 20-parsec
survey, and we are currently pursuing 
Finding the missing systems, however, will only mark the 
completion of stage two in compiling a reliable nearby-star census.
Approximately one-third of the current sample ($\sim 350$ stars)
lack accurate trigonometric parallax measurements, 
while even more stars lack close scrutiny
for either spectroscopic or resolved lower-luminosity companions. 
Those observations are beyond the scope of our current NStars project, but
are essential if we are to establish a complete catalogue of
stars and brown dwarfs within 20 parsecs of the Sun. 

\acknowledgements 
The NStars research described in this paper was 
supported partially by a grant awarded as part of the NASA Space 
Interferometry Mission Science Program, administered by the Jet Propulsion Laboratory, Pasadena.
 KLC acknowledges support from an NSF Graduate Research Fellowship.
This publication makes use of data products from the Two Micron All Sky Survey, which is
a joint project of the University of Massachusetts and the Infrared Processing and Analysis 
Center/California Institute of Technology, funded by the National Aerospace and Space 
Administration
and the National Science Foundation. We acknowledge use of the NASA/IPAC Infrared Source 
Archive (IRSA), 
which is operated by the Jet Propulsion Laboratory, California Institute of Technology, 
under contract with the  National Aerospeace and Space Administration.
We also acknowledge making extensive use of the SIMBAD database, maintained by Strasbourg 
Observatory,
and of the ADS bibliographic service. \\
This project has profited from extensive allocations of telescope time at both Kitt Peak
Observatory and Cerro-Tololo Interamerican Observatory. We thank the NOAO Telescope Allocation Committees for their support of this project 
and acknowledge John Glaspey, Darryl Willmarth, Diane Harmer, Bill Gillespie, 
Hillary Mathis, Ed Eastburn and Hal Halbedel at KPNO; Alberto Alvarez, Edgardo Cosgrove, 
Arturo Gomez, Angel Guerra, Daniel Maturana, Sergio Pizarro and
Patricio Ugarte  at CTIO.

\begin{figure}
\plotone{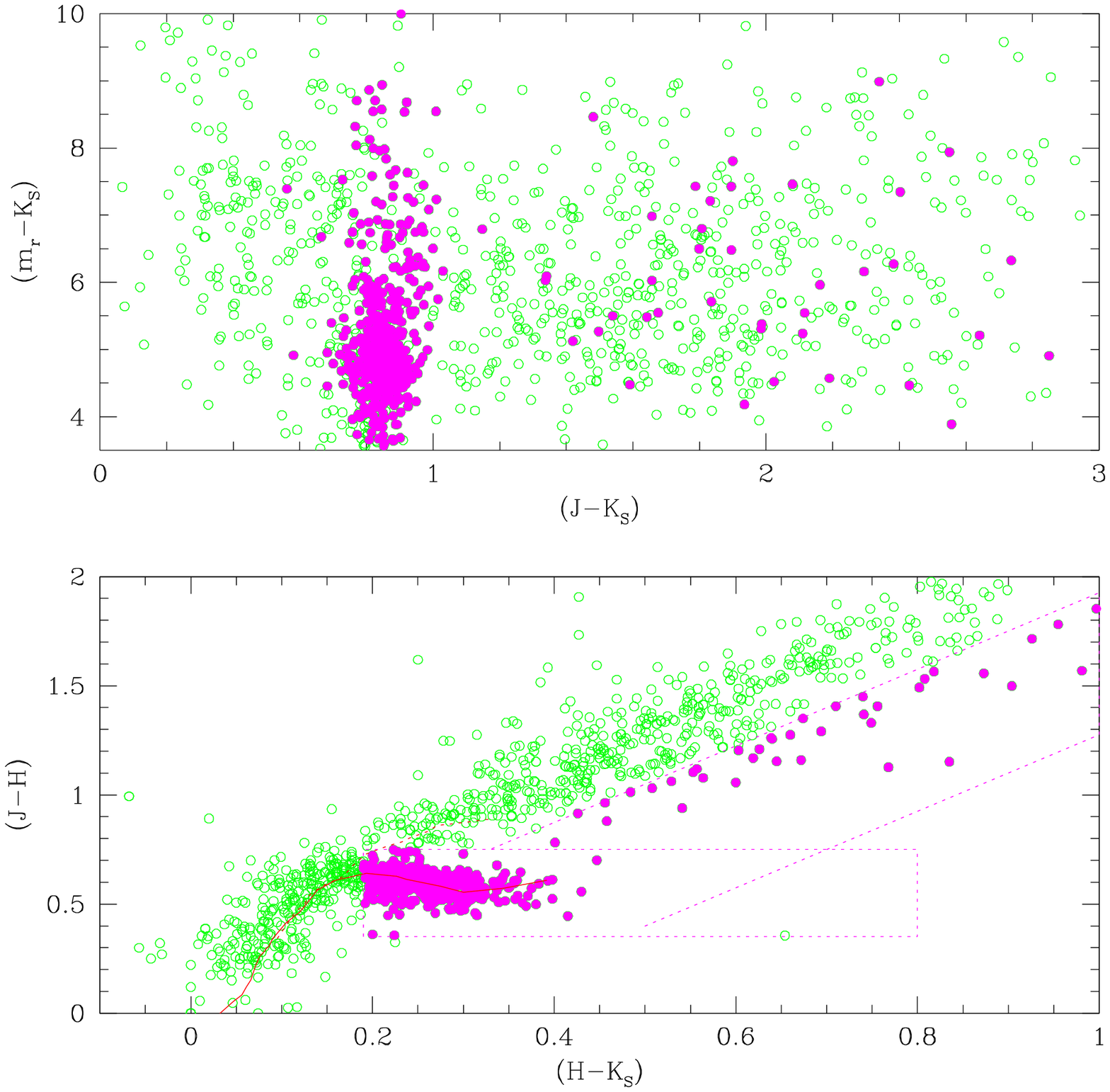}
\caption{ Optical and near-infrared colour-colour diagrams for the 
1468 NLTT2 nearby star candidates. The dotted lines in the JHK$_S$ diagram
outline the schematic M-dwarf (horizontal region) and L-dwarf r\'egimes, and
sources with dwarf-like (J-H)/(H-K$_S$) colours are plotted as
solid points in both diagrams. All sources that fall in the L-dwarf 
segment in the JHK$_S$ plane have (J-K$_S) > 1.2$ and implausible 
(m$_r$-K$_S$) colours; these are likely to be background red giants and  mismatched NLTT/2MASS pairings.}
\end{figure}
\clearpage

\begin{figure}
\caption{ Optical/near-infrared colour-colour diagrams for 1908
nearby-star candidates selected from the rNLTT. 
We show the mean relation for dwarfs (solid line) and giants (dotted line) in 
the JHK$_S$ plane. While most candidates have self-consistent
optical/near-infrared colours, there are a number of outliers. We have 
used the linear relations plotted in the upper two panels to
pick out the most extreme examples, 
identified as solid points. Those stars are listed in Appendix I.}
\end{figure}
\clearpage

\begin{figure}
\plotone{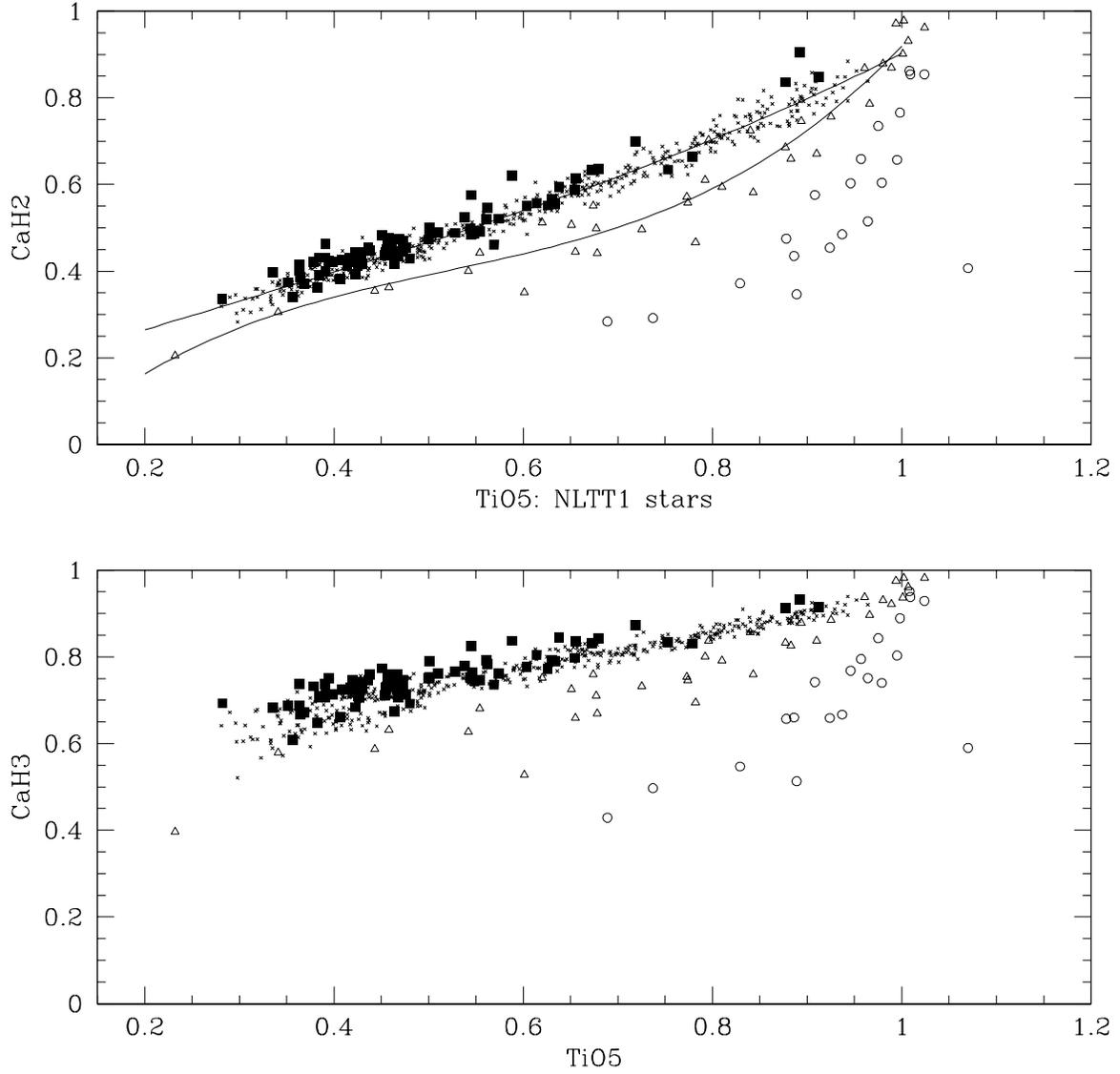}
\caption{ TiO and CaH bandstrengths of the NLTT1 stars listed in Tables 4 and 5.
The program stars are plotted as solid squares. Reference data for disk
dwarfs (plotted as ccrosses) are taken from the PMSU survey (PMSU1, PMSU2), 
while measurements of intermediate (open triangles) and extreme 
(open circles) subdwarfs are from Gizis (1997).
The solid lines in the CaH2/TiO5 diagram mark the mean relations for disk
dwarfs and intermediate subdwarfs. }
\end{figure}

\begin{figure}
\plotone{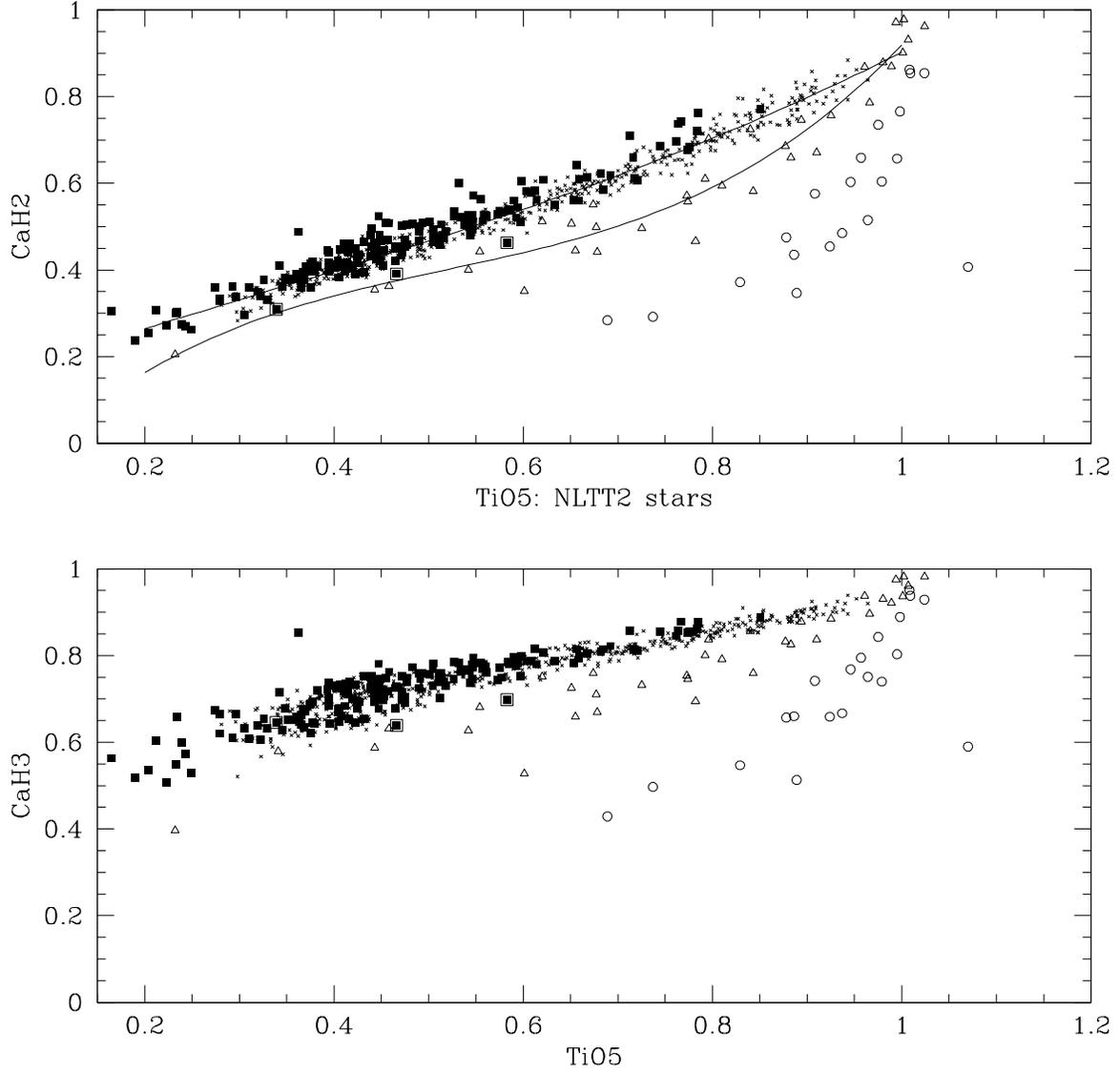}
\caption{ Relative TiO/CaH bandstrengths of the NLTT2 stars listed in Tables 4 and 5.
The symbols have the same meaning as in figure 3. The three potential 
intermediate subdwarfs mentioned
in the text (LP 16-197, LP 466-156 and LP 881-272) are identified by the points
enclosed in open squares.}
\end{figure}

\begin{figure}
\plotone{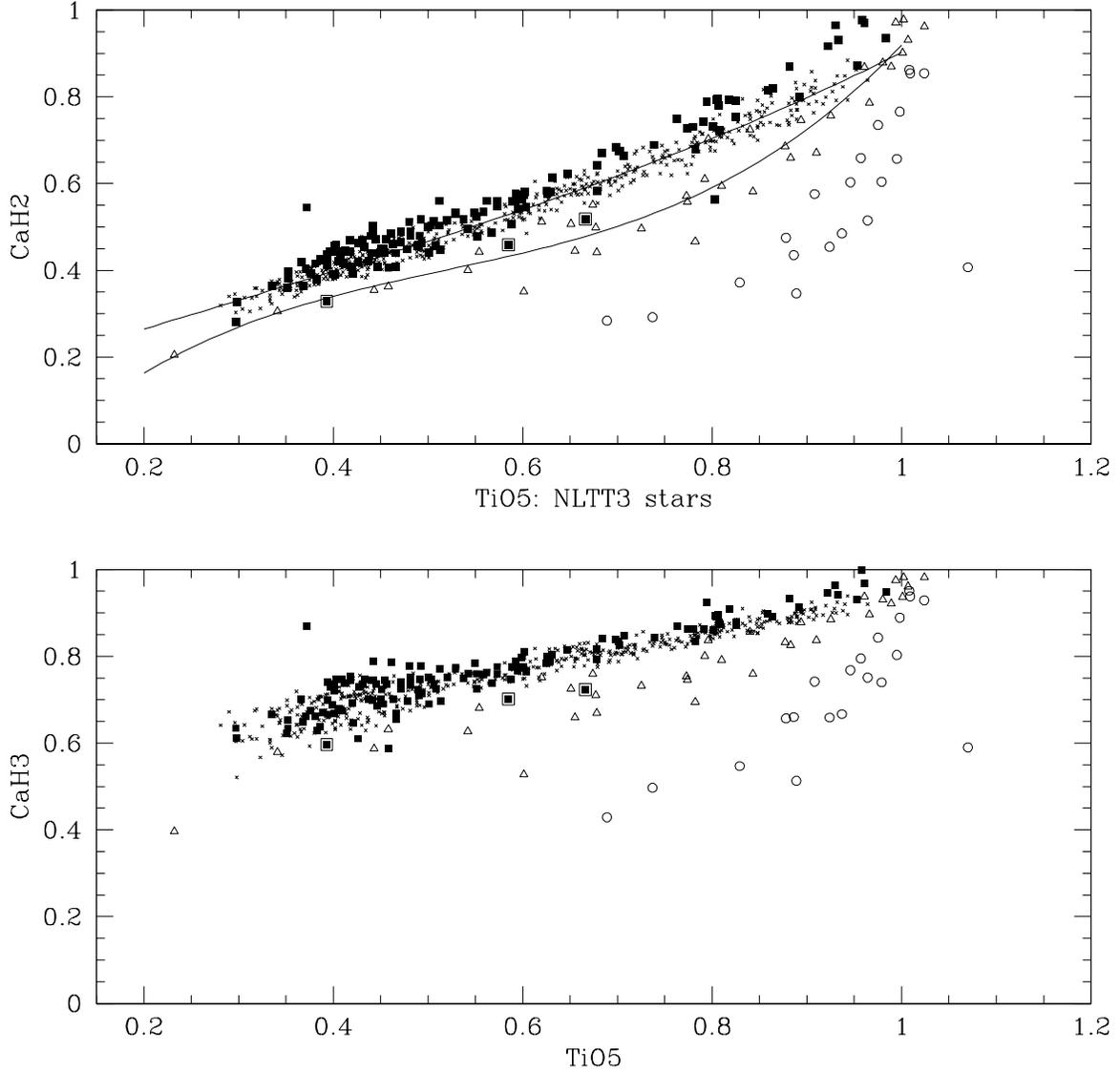}
\caption{ Relative TiO/CaH bandstrengths of the NLTT3 stars listed in 
Tables 4 and 5. This dataset includes a higher proportion of early-type 
M dwarfs than the NLTT1 and NLTT2 samples. The three mildly metal-poor stars mentioned
in the text (LP 97-817, G 174-25 and LP 109-57) are identified by the points
enclosed in open squares. The symbols have the same meaning as in figures 3 and 4.}
\end{figure}

\begin{figure}
\figurenum{6a}
\plotone{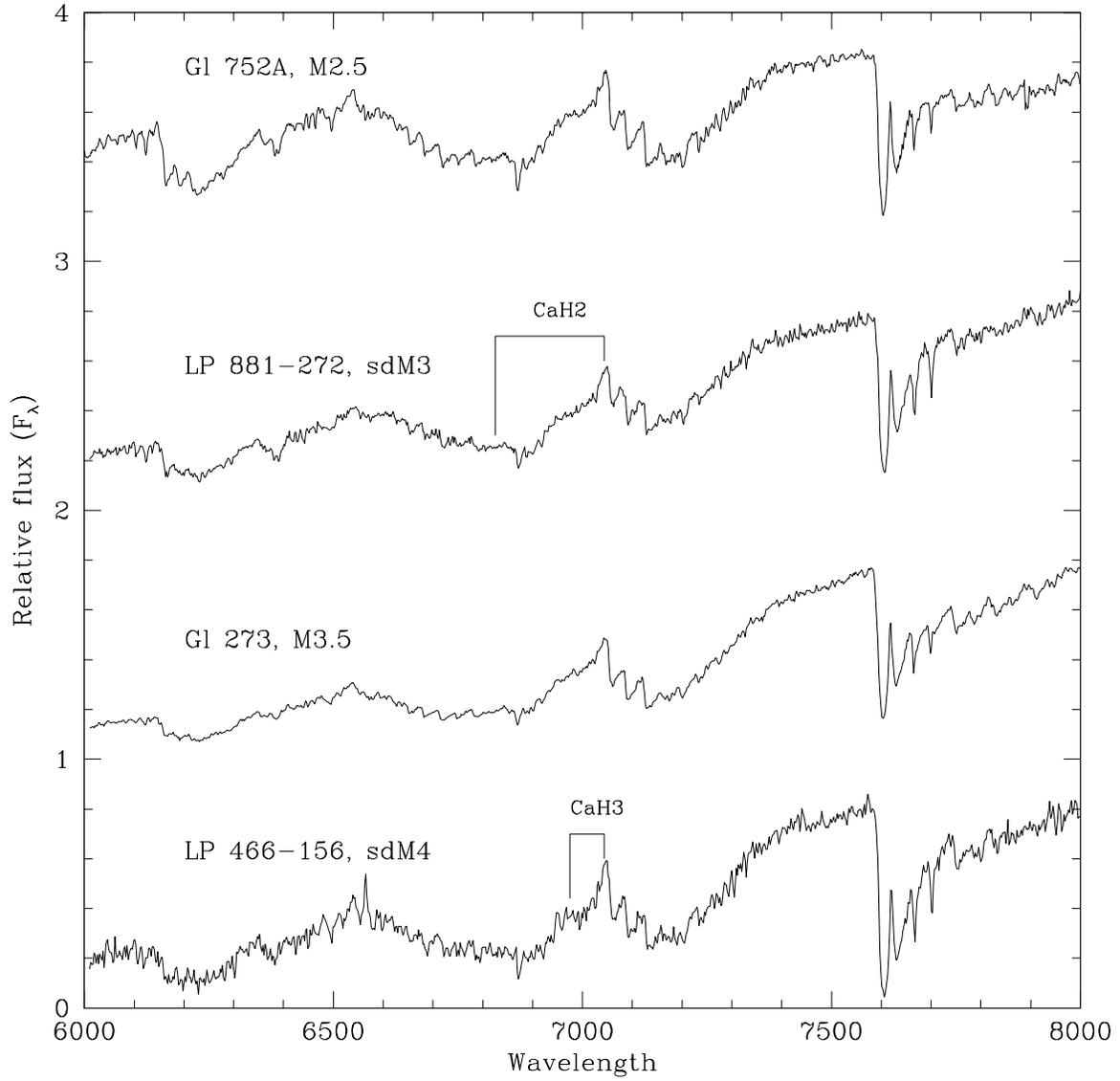}
\caption{ LP 466-156 and LP 881-272, candidate intermediate
 subdwarfs from the NLTT2 sample. The locations of the CaH2 and CaH3 in-band
and pseudo-continuum measurements are shown; the latter is centred on the peak
flux immediately shortward of the $\lambda$7050\AA\ TiO band.}
\end{figure}

\begin{figure}
\figurenum{6b}
\plotone{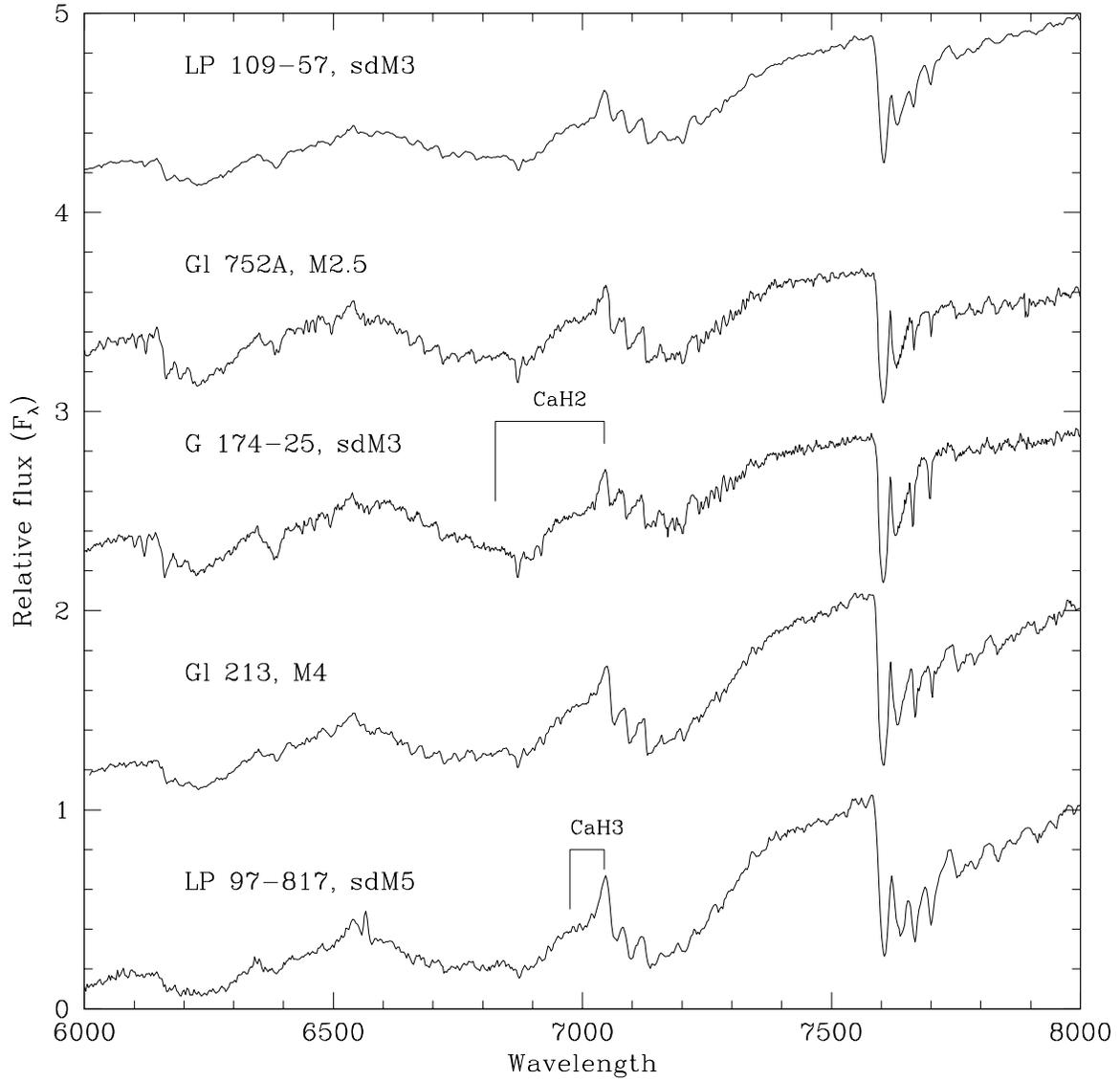}
\caption{ LP 109-57, G 174-25 and LP 97-817, a further three 
candidate intermediate subdwarfs from the NLTT3 sample. We again identify the
location of the reference points for the CaH2 and CaH3 band indices.}
\end{figure}

\begin{figure}
\figurenum{7}
\plotone{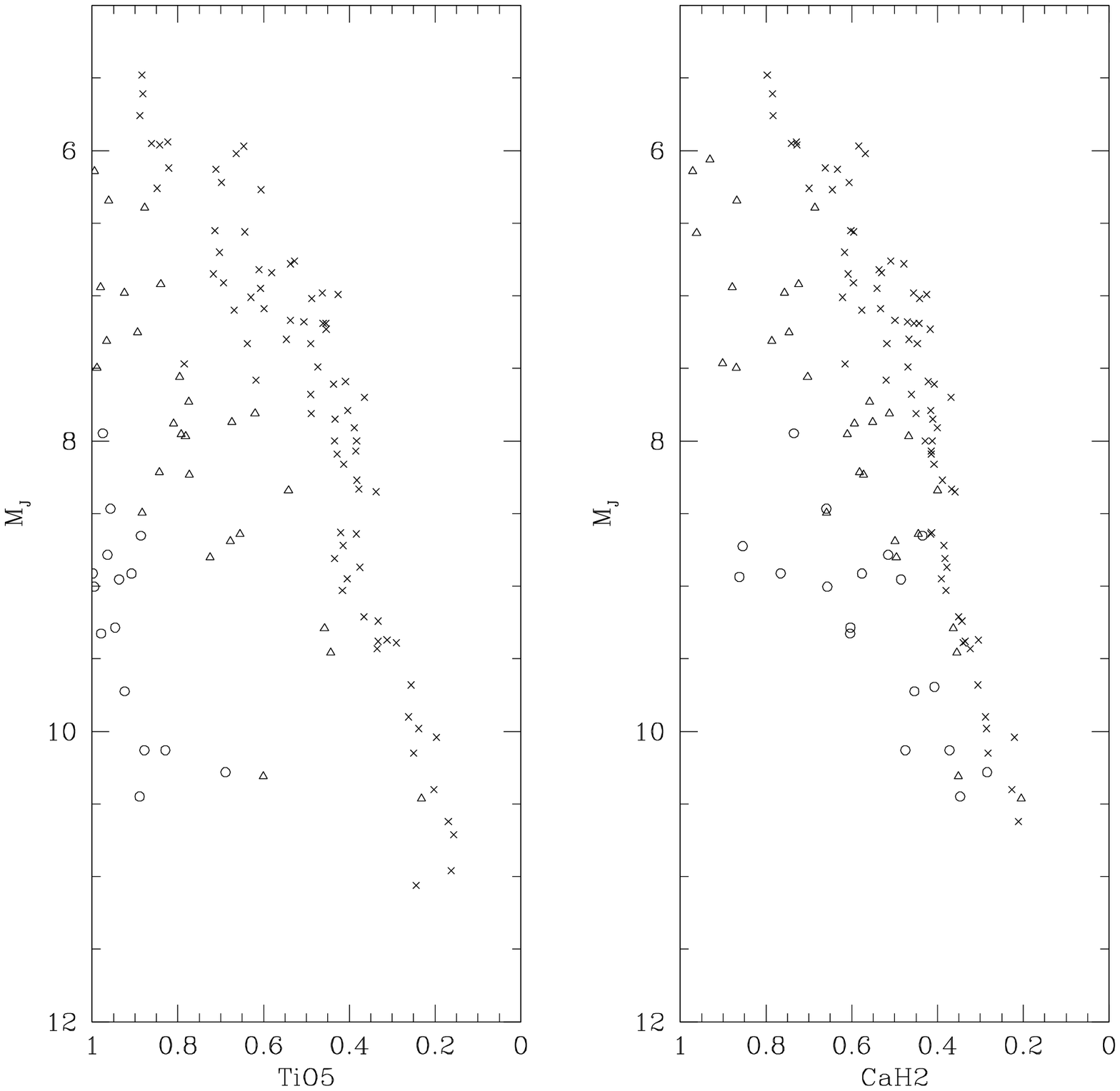}
\caption{ The (M$_J$, TiO5) and (M$_J$, CaH2) H-R diagrams for late-type
dwarfs and subdwarfs. The crosses mark data for single stars with accurate
trigonometric parallax measurements (${\sigma_\pi \over \pi} < 7\%$) from\
the 8-parsec sample. sdM dwarfs are plotted as open triangles, and 
esdM dwarfs as open circles.}
\end{figure}

\begin{figure}
\figurenum{8}
\plotone{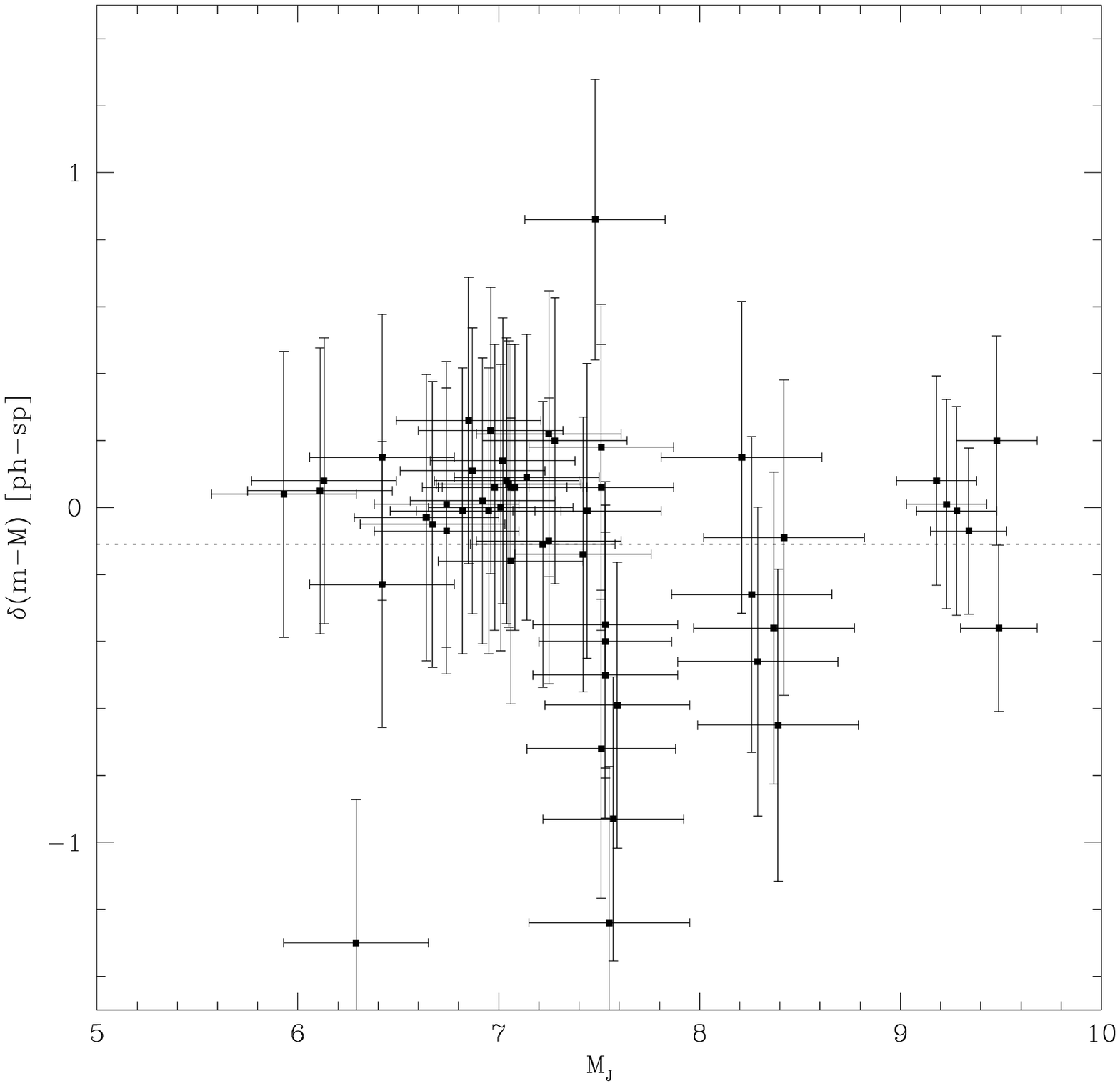}
\caption{ A comparison between photometric and spectrophotometric distance
 moduli for stars from the NLTT2 sample included in the present paper.
The dotted line marks the formal mean difference (-0.13 mag). 
The scatter ($\sigma=0.36$ mag) is consistent
with the formal uncertainties in the individual measurements.}
\end{figure}

\begin{figure}
\figurenum{9}
\plotone{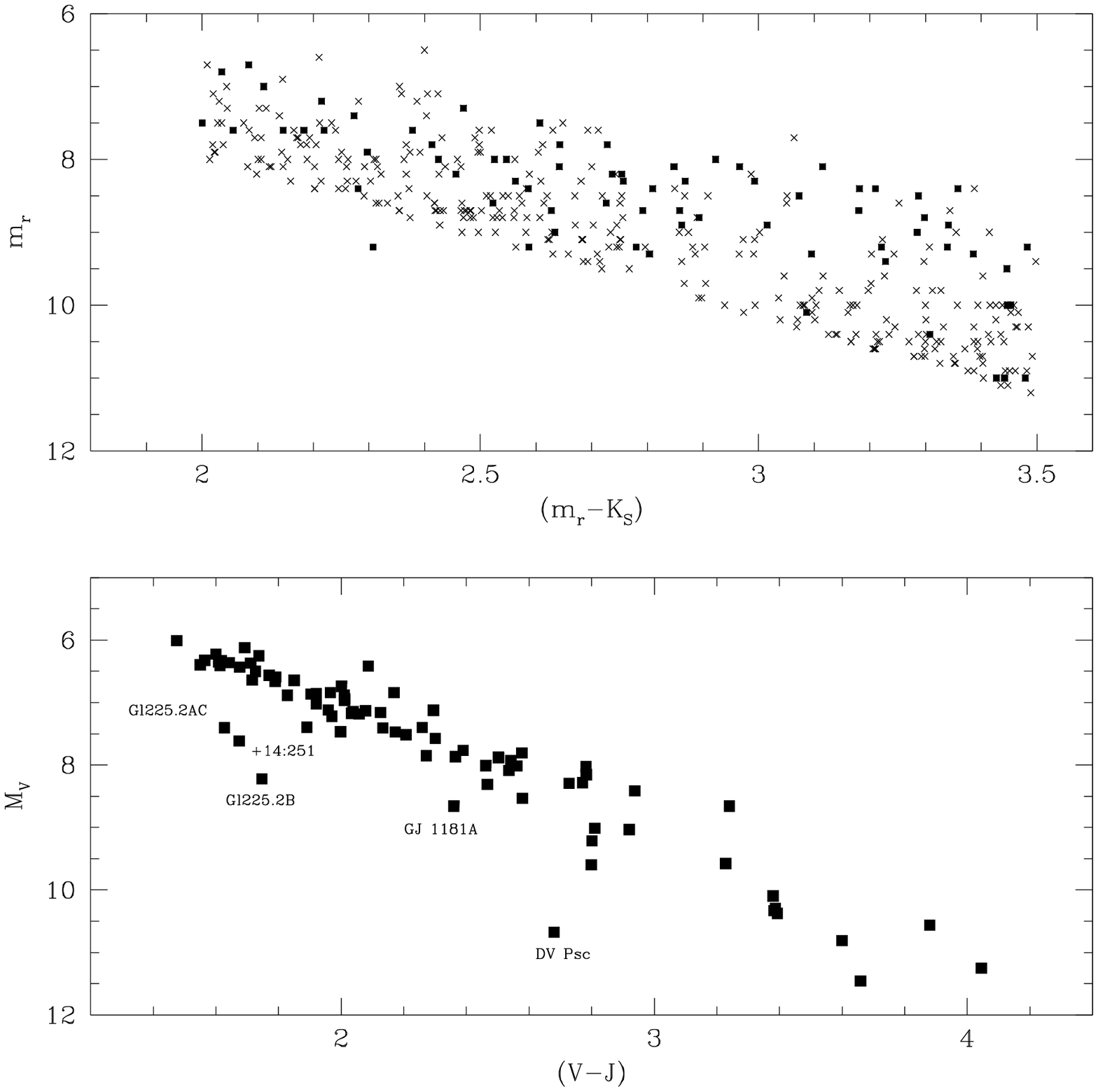}
\caption{ Photometric data for K and early-type M-dwarf nearby-star
candidates selected from the rNLTT. 
The upper panel plots the (m$_r$, (m$_r-K_S$)) distribution. Stars with
distances measureed at less than 20 parsecs are plotted as solid points. The 
lower diagram plots the (M$_V$, (V-J)) distribution of the latter stars.
The labeled points are discussed in the text; with the exception of GJ 1181A, 
the distances to these stars are probably underestimated. }
\end{figure}

\begin{figure}
\figurenum{10}
\plotone{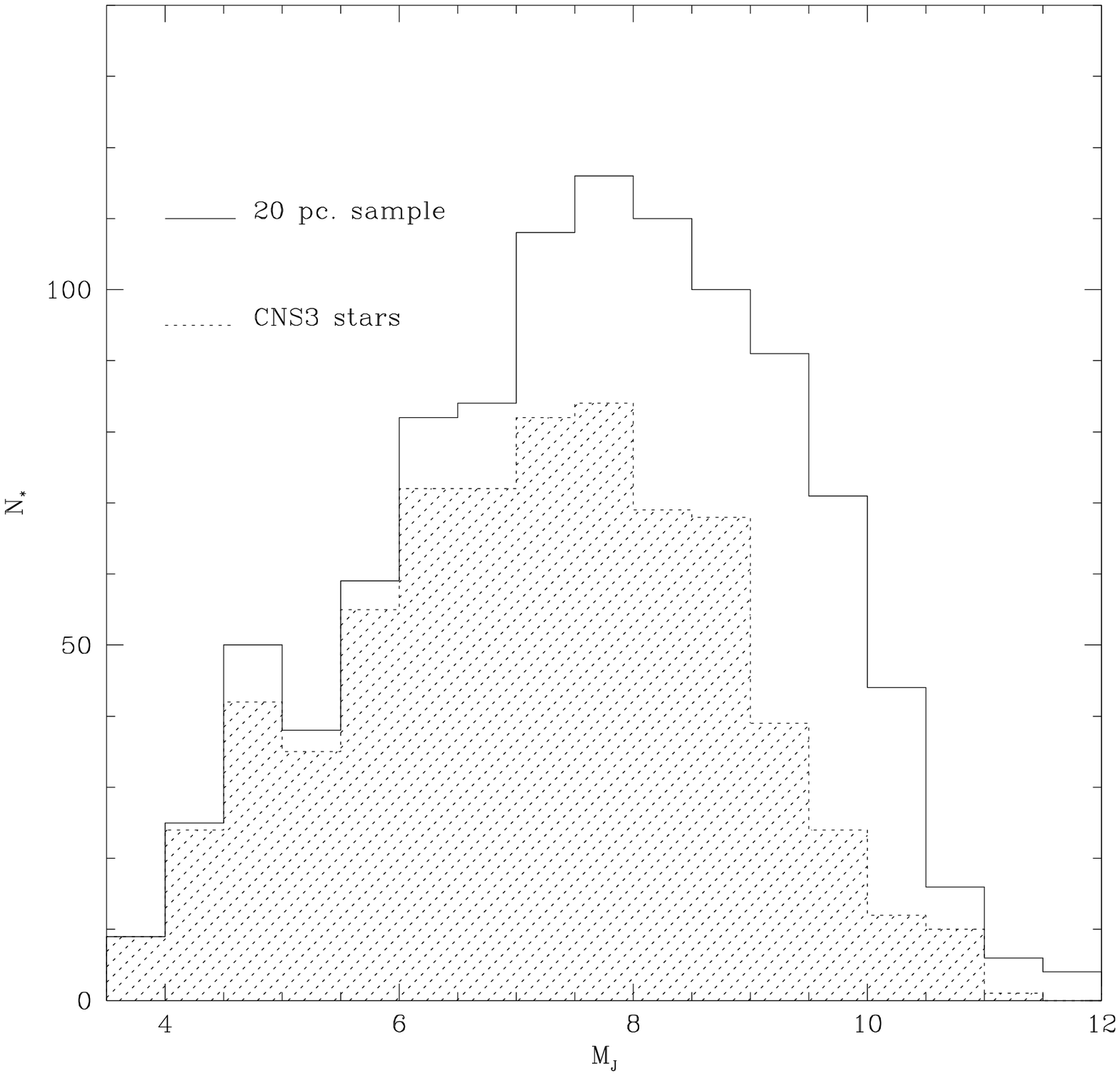}
\caption { Statistics of the 20-parsec census: we plot the
number-magnitude distribution for all stars identified as within 20
parsecs of the Sun, with 
the shaded area marking the contribution from stars catalogued in the CNS3.
Most of the additions lie at faint magnitudes, M$_J > 7$.}
\end{figure}

\begin{figure}
\figurenum{11}
\plotone{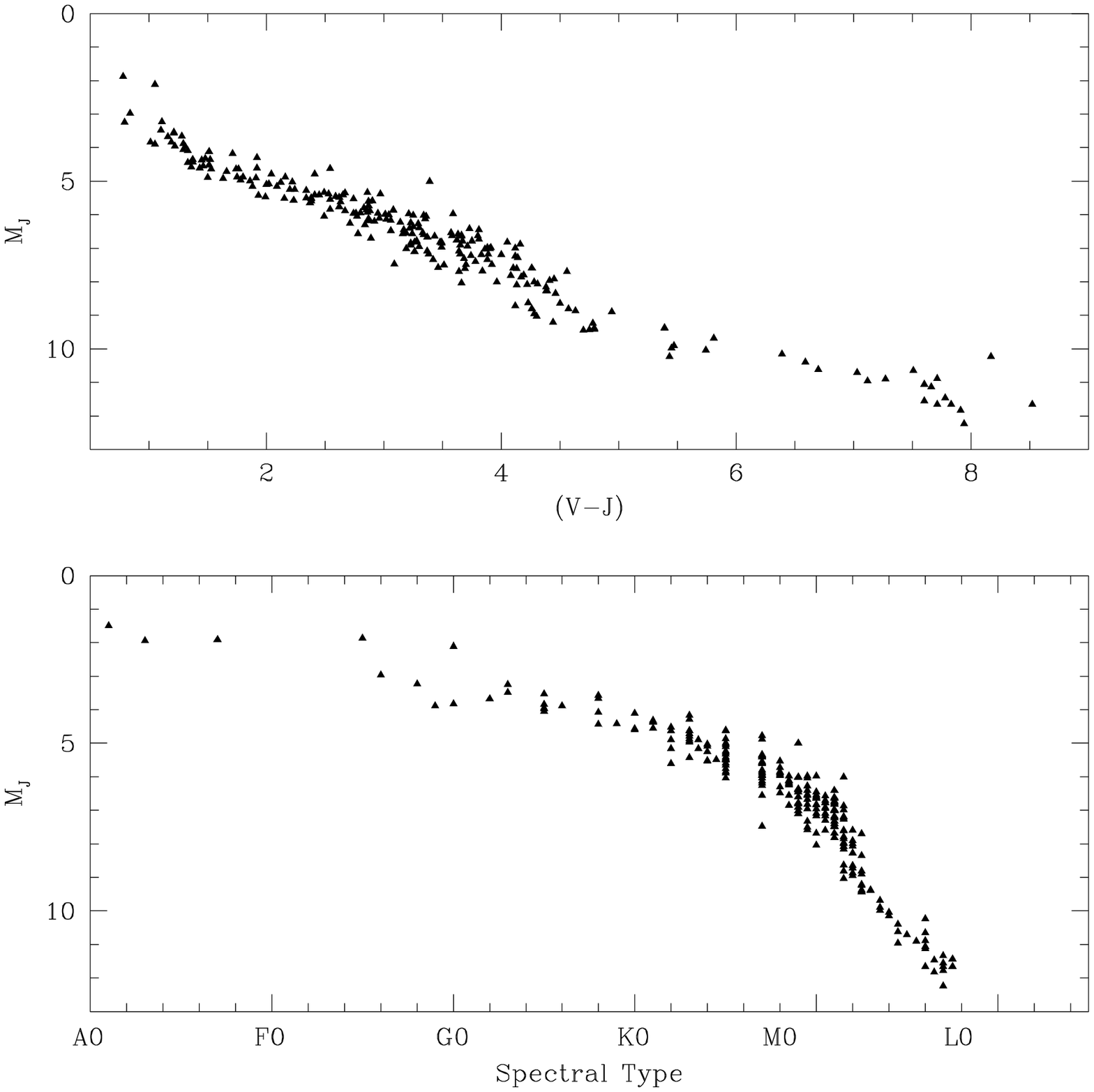}
\caption{ The (M$_J$, (V-J)) and (M$_J$, sp. type) relations, defined by stars
with reliable photometry and trigonometric parallaxes measured to an accuracy
${\sigma_\pi \over \pi} < 7\%$. Note the significant change in slope in
the main sequence at $4 < (V-J) < 4.5$, corresponding to spectral types M2/M3.}
\end{figure}

\begin{figure}
\figurenum{12}
\plotone{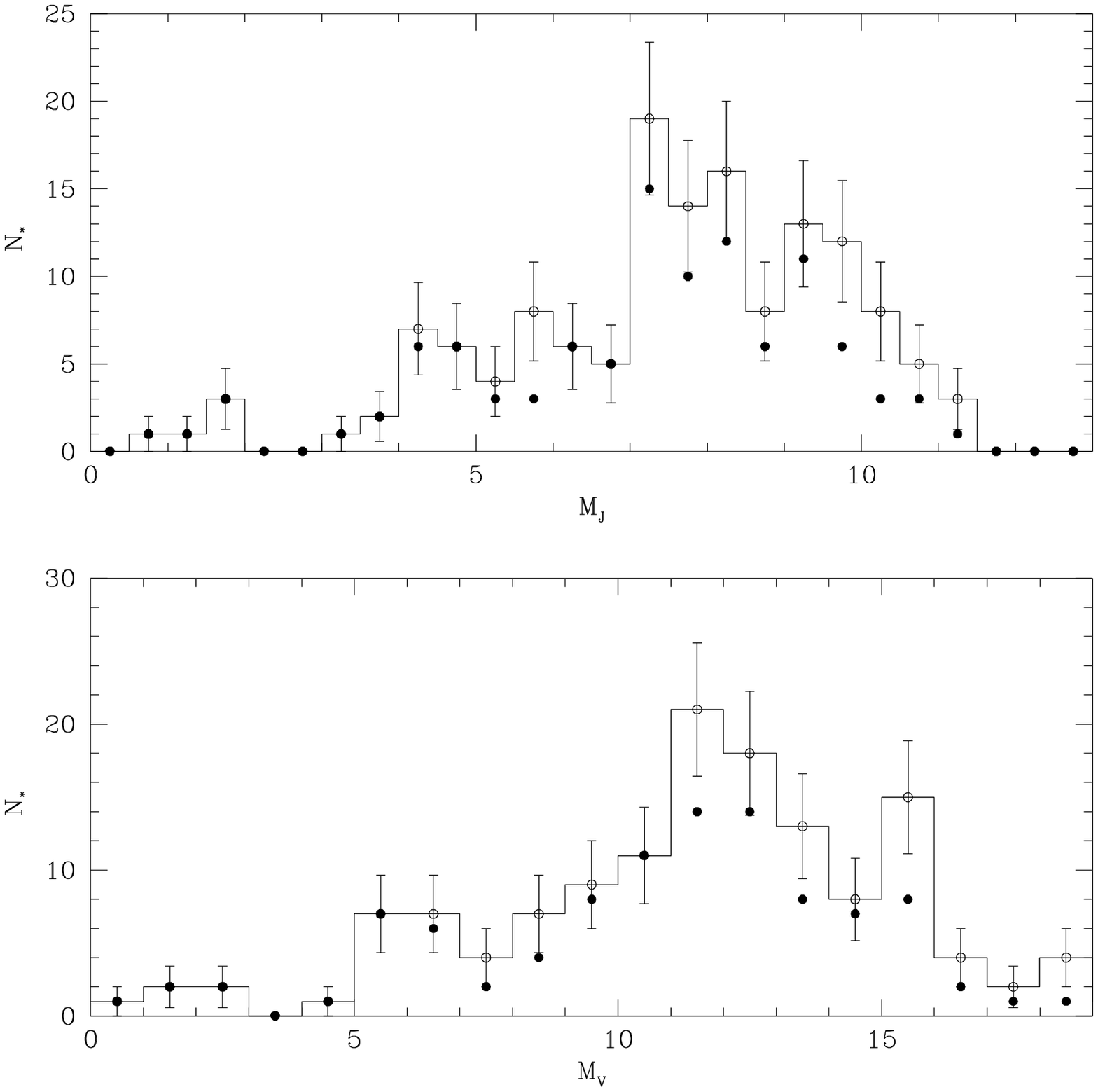}
\caption{ A comparison between the V-band and J-band luminosity functions
derived from the northern 8-parsec sample. The 
histograms (and open circles) plot data for all stars, with 
the error bars showing the formal Poisson uncertainties; the solid 
points mark the contribution from  single stars and primaries in multiple systems.}
\end{figure}

\begin{figure}
\figurenum{13}
\plotone{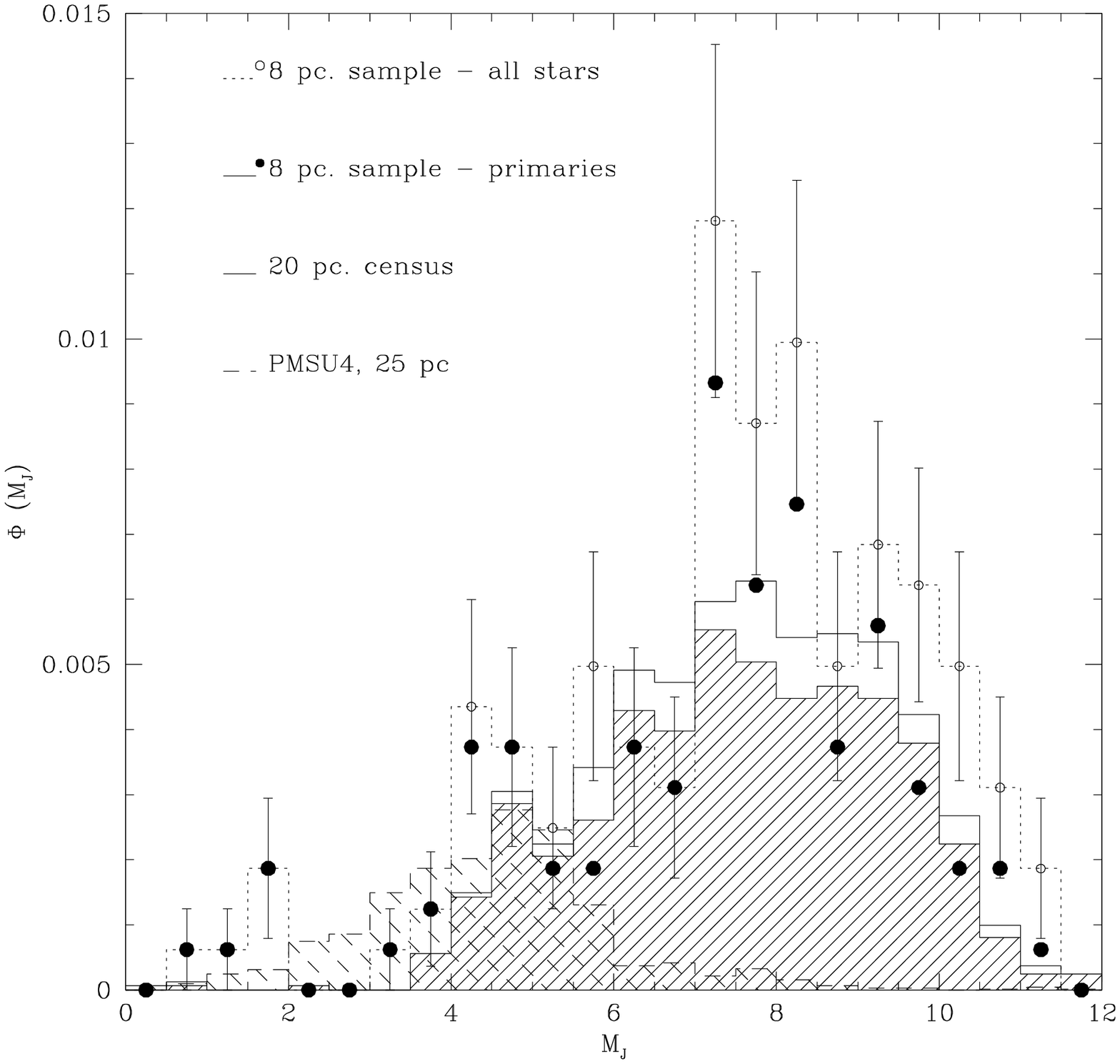}
\caption{ The J-band luminosity function: the space densities derived
from the 20-parsec census are plotted as a solid-line histogram. The 
contribution from single stars and primaries is indicated by the
bend-dexter hatched histogram (solid lines). These results can be
compared against data for the PMSU4 sample at bright magnitudes 
(dashed line, bend-sinister shaded histogram), and for the 
northern 8-parsec sample (open circles and dotted histogram).
As in Figure 12, the solid points mark the number densities
of single stars and primaries in the latter sample. Malmquist and 
Lutz-Kelker corrections, as appropriate, have been applied in 
constructing these luminosity functions.}
\end{figure}

\begin{figure}
\figurenum{14}
\plotone{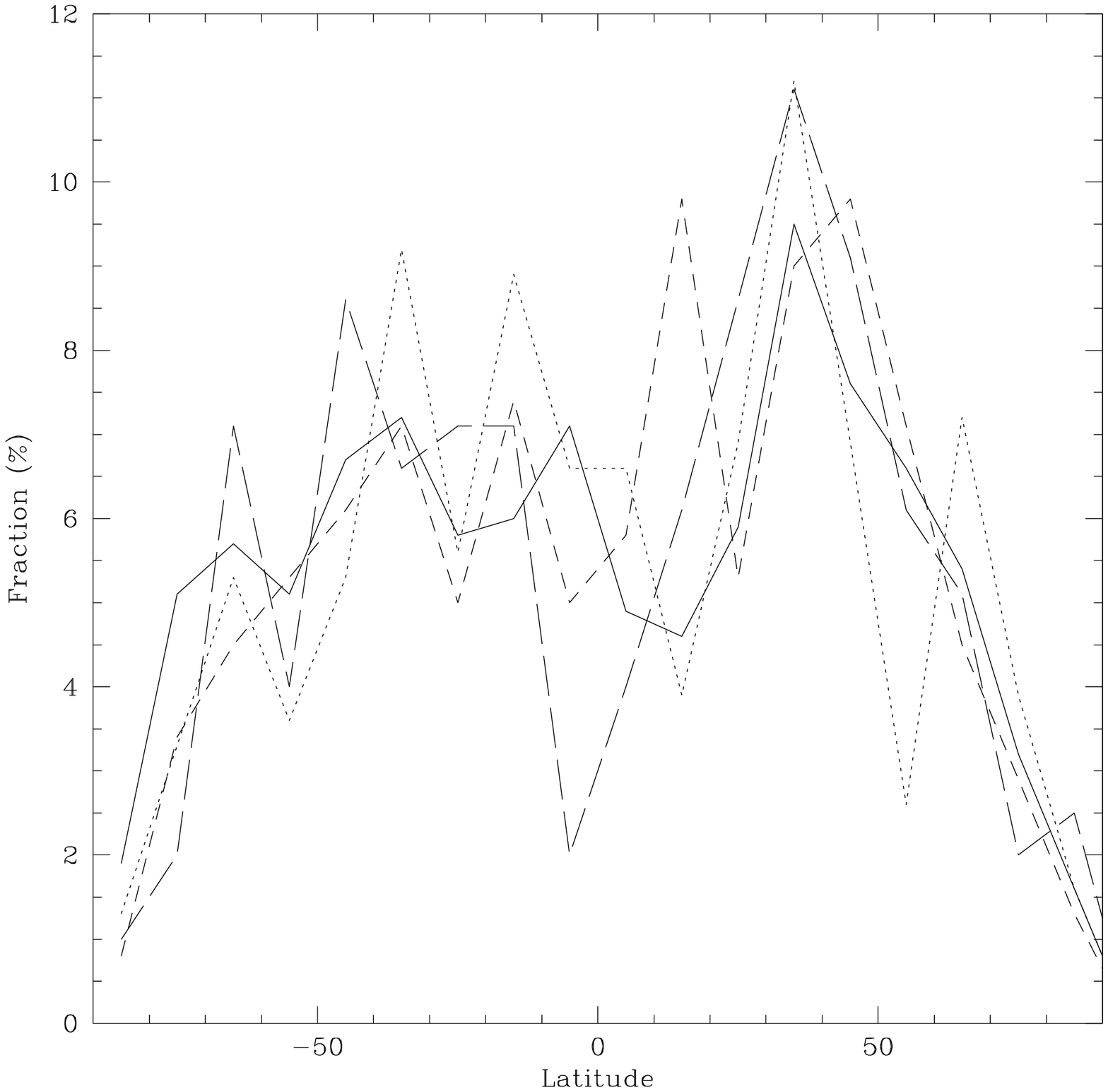}
\caption { The latitude distribution of stars in the 20-parsec sample.
We plot the percentage of stars in 10-degree bins in latitude for 
three absolute magnitude intervals: M$_J < 7$ (dotted line); $7 < M_J < 9$
(dashed line); and M$_J > 9$ (long-dashed line). The solid line plots
similar data for all 2M2nd NLTT dwarfs with $8 < m_r < 10$. There is evidence
for a moderate deficit in the Plane at the faintest luminosities. }
\end{figure}

\clearpage
\include{Reid.tab1}

\include {Reid.tab2}

\include {Reid.tab3}

\include {Reid.tab4}

\include {Reid.tab5}

\include {Reid.tab6}

\include {Reid.tab7}

\include {Reid.tab8}

\include {Reid.tab9}

\include {Reid.apptab1}

\include {Reid.apptab2}
\end{document}

%% file: Reid.tab1.tex
\begin{deluxetable}{rrrrrrrrrccccccccccc}
\rotate
\tabletypesize{\scriptsize}
\tablecolumns{18}
\tablenum{1}
\tablewidth{0pt}
\tablecaption{Known nearby stars from NLTT2}
\tablehead{\colhead{name} & \colhead{ Other} & 
\colhead{ $\alpha$ (2000)} & \colhead{$\delta$} & \colhead{$m_r$} &
\colhead {J} &  \colhead {H} &\colhead {K$_S$} & \colhead{V} & \colhead{B-V} & \colhead{V-R} & \colhead{V-I} &
\colhead{src.} & \colhead{$\pi_{trig}$} & \colhead{src.} & \colhead {Sp. type} & \colhead{ (m-M) } &  
\colhead{M$_J$} & \colhead {Notes }}
\startdata
 G 130-53$^*$& LHS 1037  &  0:11:56.5& 33:03:17&  12.4&    9.04&    8.40&    8.12&  13.00&   1.62&   1.12&   2.56&  1& \nodata&& \nodata&   1.55$\pm$0.30&   7.49&  \\
   86- 67& LHS 1049   &  0:15:43.4&-67:59:35&  13.8&    8.85&    8.24&    7.96&  12.50&  \nodata&  1.07&  2.39&  10& \nodata&&M2.5&   1.77$\pm$0.23&   7.08&  \\
   86- 66$^*$& LHS 1051   &  0:15:51.4&-67:59:51&  11.6&    7.84&    7.21&    6.93&  10.95&  \nodata&  &  &  10& \nodata&&M0.5&   1.52$\pm$0.23&   6.32&  \\
  404- 80   &  G 131-50  &  0:17:58.6& 20:57:19&  12.0&    8.67&    8.03&    7.83&  11.76&  \nodata&   0.94&   2.00&  1& \nodata&& \nodata&   2.42$\pm$0.30&   6.25& 1  \\
  881- 64$^{**}$& GJ 2005    &  0:24:44.2&-27:08:25&  14.4&    9.26&    8.53&    8.23&  15.42&  \nodata&   1.84&   4.02&  7& 133$\pm$9&2&M5.5&  -0.69$\pm$0.13&   9.95& triple \\
  825-288       & G 266-123  &  0:29:58.6&-24:38:26&  12.9&    9.21&    8.55&    8.26&  12.63&   1.47&  \nodata&  \nodata&  8& \nodata&& \nodata&   2.28$\pm$0.41&   6.93&  \\
 G 172-11$^*$      &            &  0:35:53.6& 52:41:36&  14.6&   10.65&   10.10&    9.83&  15.10&  \nodata&   1.27&   2.91&  1& 62$\pm$9&7&\nodata&   1.51$\pm$0.23&   9.14&  \\
  -34:239$^{**}$& GJ 1016    &  0:41:30.5&-33:37:31&  10.8&    7.70&    7.08&    6.86&  10.58&   1.43&  \nodata&  \nodata&  2&  50$\pm$2&1&K7.0&   1.49$\pm$0.08&   6.21&  \\
\enddata
\tablecomments{ \\
$^{**}$ identifies stars with formal distance estimates $d \le 20$ parsecs  \\
$^*$ identifies stars with formal distance estimates within $1\sigma$ of 20 parsecs  \\
Column 1 lists the designation from the NLTT catalogue: R=Ross, W=Wolf, Oxf=Oxford catalogue.  We
have added Lowell Observatory Proper Motion Survey identifications (Giclas {\sl et al.}, 1971) where appropriate. 
Column 2 lists an alternative name, usually from either the CNS3 or the LHS catalogue; \\
Columns 3 and 4 list the position of the 2MASS source; Column 5 gives the Luyten m$_r$ magnitude; 
Columns 6-8 list the 2MASS photometry; \\
Columns 9-12 list the optical photometry, and column 13 gives the source:
1 - Weis (1984, 1986, 1987, 1988, 1991a, 1991b, 1996); 
2 - Hipparcos (ESA, 1997); 3 - CNS3; 4 - Reid (1990); 5 - Bessell (1990); 6 - Leggett (1992); 
7 - Patterson {\sl et al.} (1998); 8 - Ryan (1989, 1992); 9 - Caldwell {\sl et al.} (1984); 
10 - Eggen (1987); 11 - Fleming (1998); 12 - Laing (1989); 13 -  USNO, Harrington {\sl et al.} 
(1993) and refs within; 14 - Dawson \& Forbes (1989, 1992); 15 - Sandage \& Kowal (1986); 
16 - Norris {\sl et al.}, (1999); 17 - van Altena, Lee \& Hoffleit (2001). \\
Column 14 lists the trigonometric parallax and uncertainty; column 15 gives the source: \\
1 - Hipparcos (ESA, 1997); 2 - Ianna \& Bessell (1986); 3 - Harrington \& Dahn (1980); 
4 - CNS3; 5 - Dahn {\sl et al.} (1988); 6 - Oppenheimer {\sl et al.} (2001); 
van Altena, Lee \& Hoffleit (2001).\\
Column 16 gives the estimated distance modulus and column 17 the
corresponding M$_J$. \\
Notes: 
1 - cpm G 131-51; 2 - cpm LP 193-584; 3 - cpm G172-34; 4 - binary, Jao {\sl et al.} (2003);
5 - cpm HD 263175; 6 - cpm G 40-7 }
\end{deluxetable}

%% file: Reid.tab2.tex
\begin{deluxetable}{rrrrrrrrrccccccccccc}
\rotate
\tabletypesize{\scriptsize}
\tablecolumns{18}
\tablenum{2}
\tablewidth{0pt}
\tablecaption{Known nearby stars from NLTT3}
\tablehead{\colhead{name} & \colhead{ Other} & 
\colhead{ $\alpha$ (2000)} & \colhead{$\delta$} & \colhead{$m_r$} &
\colhead {J} &  \colhead {H} &\colhead {K$_S$} & \colhead{V} & \colhead{B-V} & \colhead{V-R} & \colhead{V-I} &
\colhead{src.} & \colhead{$\pi_{trig}$} & \colhead{src.} & \colhead {Sp. type} & \colhead{ (m-M) } &  
\colhead{M$_J$} & \colhead {Notes }}
\startdata
    G 217-40     &            &  0:15:14.6& 53:04:45&  14.6&   10.80&   10.21&    9.97&  14.39&  \nodata&   1.04&   2.38&  1& \nodata&  & \nodata&   3.91$\pm$0.30&   6.89&  \\
 404- 81        &   G 131-50  &  0:17:59.2& 20:57:24&  11.0&    8.24&    7.57&    7.41&  11.08&  \nodata&   0.88&   1.79&  1& \nodata&  & \nodata&   2.30$\pm$0.30&   5.94&  \\
 585- 86$^{**}$&  BRI0021   &  0:24:24.6& -1:58:19&  19.0&   12.02&   11.08&   10.54&  19.42&  \nodata&   2.20&   4.46& 10& 83$\pm$3&  4& M9.5&   0.42$\pm$0.09&  11.60&  \\
   W   4        &  Gl 22.1   &  0:33:05.6& 42:00:05&  11.3&    8.31&    7.63&    7.49&  10.94&   1.31&  \nodata&  \nodata&  2& 29$\pm$2&  1& M0.0&   2.64$\pm$0.16&   5.67&  \\
 350- 20$^{**}$& LTT10301A  &  0:50:33.2& 24:49:00&  11.7&    7.92&    7.33&    7.09&  12.09&  \nodata&   1.18&   2.72&  1&  83$\pm$9&  1& M3.5&   0.33$\pm$0.16&   7.59&  \\
 406- 76$^{**}$& GJ 1026 AB &  1:03:14.2& 20:05:52&  11.3&    7.65&    7.09&    6.82&  11.40&   1.61&   1.09&   2.43&  1&  61$\pm$5&  1& M2.0&   0.81$\pm$0.15&   6.84&  \\
 G 132-51        &            &  1:03:42.1& 40:51:15&  12.7&    9.18&    8.72&    8.48&  13.01&  \nodata&   1.10&   2.51&  1& \nodata&  & \nodata&   2.07$\pm$0.30&   7.11&  \\
 707- 58$^*$& Gl  56.1   &  1:18:16.0&-12:53:59&  11.6&    8.37&    7.74&    7.55&  11.79&   1.47&   1.00&   2.22& 8&  45$\pm$4&  1& M2.0&   1.63$\pm$0.14&   6.74&  \\
\enddata
\tablecomments{ 
$^{**}$ identifies stars with formal distance estimates $d \le 20$ parsecs  \\
$^*$ identifies stars with formal distance estimates within $1\sigma$ of 20 parsecs  \\
Column 1 lists the designation from the NLTT catalogue: R=Ross, W=Wolf, Oxf=Oxford catalogue.  We
have added Lowell Observatory Proper Motion Survey identifications (Giclas {\sl et al.}, 1971) where appropriate. 
Column 2 lists an alternative name, usually from either the CNS3 or the LHS catalogue; \\
Columns 3 and 4 list the position of the 2MASS source; Column 5 gives the Luyten m$_r$ magnitude; 
Columns 6-8 list the 2MASS photometry; \\
Columns 9-12 list the optical photometry, and column 13 gives the source:
1 - Weis (1984, 1986, 1987, 1988, 1991a, 1991b, 1996); 
2 - Hipparcos (ESA, 1997); 3 - CNS3; 4 - Sandage \& Kowal (1986); 5 - Ryan (1989, 1990); 6 - Leggett (1992); 7  - Bessell (1990); 9 - Koen \etal, (2002);
9 - Phan-Bao \etal, (2001); 10 - Dahn \etall (2002);  
11 - Stauffer \etall (1991); 12 - Fleming (1998). \\
Column 14 lists the trigonometric parallax and uncertainty; column 15 gives the source: \\
1 - Hipparcos (ESA, 1997); 2 - CNS3; 3 - Dahn \etal, (2002); 4 - van Altena, Lee \& Hoffleit (2001)\\
Column 16 gives the estimated distance modulus and column 17 the corresponding
value of M$_J$.
\\ Notes: 
1 - LP 320-16/G 148-48: Joint optical photometry, JHK$_S$ data taken from All-Sky release. \\
}
\end{deluxetable}

%% file: Reid.tab3.tex
\begin{deluxetable}{rrrrrrrrrccccccccccc}
\rotate
\tabletypesize{\scriptsize}
\tablecolumns{15}
\tablenum{3}
\tablewidth{0pt}
\tablecaption{SAAO photometry of NLTT2 and NLTT3 stars}
\tablehead{\colhead{name} & \colhead{ Other} & 
\colhead{ $\alpha$ (2000)} & \colhead{$\delta$} & \colhead{$m_r$} &
\colhead {J} &  \colhead {H} &\colhead {K$_S$} & \colhead{V} & \colhead{B-V} & \colhead{V-R} & \colhead{V-I} &
\colhead{N$_{obs}$}  & \colhead{ (m-M) } &  
\colhead{M$_J$} & \colhead {Notes }}
\startdata
 NLTT2 & \\
  644- 94   &            &  0:09:13.5& -4:08:02&  12.0&    8.60&    7.94&    7.71&  12.03&   1.51&   1.02&   2.23&  1&   1.87$\pm$0.23&   6.73&  \\
  880-803   &            &  0:16:20.1&-31:27:39&  13.3&    9.53&    8.91&    8.68&  13.01&\nodata&   1.03&   2.29&  1&   2.79$\pm$0.23&   6.74&  \\
  464-511   &            &  0:17:11.8& 14:54:52&  13.9&   10.09&    9.45&    9.23&  13.70&\nodata&   1.06&   2.37&  1&   3.16$\pm$0.30&   6.93&  \\
  86- 93    &            &  0:18:34.7&-68:46:59&  11.6&    8.58&    7.95&    7.75&  11.32&   1.44&   0.89&   1.75&  1&   2.75$\pm$0.30&   5.83&  \\
  86- 82    &            &  0:21:54.9&-68:31:25&  11.6&    8.25&    7.57&    7.35&  11.13&   1.45&   0.92&   1.87&  1&   2.20$\pm$0.23&   6.05&  \\
  881- 50  & G 267-87   &  0:22:33.9&-31:10:24&  14.5&   10.66&   10.03&    9.74&  14.27&\nodata&   1.07&   2.41&  1&   3.68$\pm$0.23&   6.98&  \\
  881-102  &            &  0:28:07.9&-26:52:28&  15.5&   10.64&    9.99&    9.79&  14.45&\nodata&   1.07&   2.44&  1&   3.42$\pm$0.23&   7.22&  \\
\enddata
\tablecomments{ 
$^{**}$ identifies stars with formal distance estimates $d \le 20$ parsecs  \\
$^*$ identifies stars with formal distance estimates within $1\sigma$ of 20 parsecs  \\
Column 1 lists the designation from the NLTT catalogue: R=Ross, W=Wolf, Oxf=Oxford catalogue.  We
have added Lowell Observatory Proper Motion Survey identifications (Giclas {\sl et al.}, 1971) where appropriate. \\
Column 2 lists an alternative name, usually from either the pCNS3 or the LHS catalogue; \\
Columns 3 and 4 list the position of the 2MASS source; Column 5 gives the Luyten m$_r$ magnitude; 
Columns 6-8 list the 2MASS photometry; \\
Columns 9-12 list the optical photometry, and column 13 gives the number of observations; \\
Column 14 gives the estimated distance modulus and column 15 the corresponding M$_J$. Notes: \\
1. probably variable; constant colours; 2. probably variable; 3. possibly affected by nearby field star; 
4. crowded field; 5. LP 767-16 is a close binary, separation $\sim1.5$ arcseconds.  Both 2MASS and the
optical data provide joint photometry; the distance estimate assumes an equal-magnitude system. \\ 
}
\end{deluxetable}

%% file: Reid.tab4.tex
\begin{deluxetable}{rrrrrrrrrc}
\tabletypesize{\scriptsize}
\tablecolumns{10}
\tablenum{4}
\tablewidth{0pt}
\tablecaption{Narrowband indices for NLTT dwarfs }
\tablehead{\colhead{name} &
\colhead{ TiO5} & \colhead{CaOH} & \colhead{CaH1} &
\colhead {CaH2} &  \colhead {CaH3} 
&\colhead {TiO-a} & \colhead{VO-a} & \colhead{H$\alpha$} & \colhead {Sp. Type }}
\startdata
NLTT1 \\
 LP 348-40   &   0.43 &   0.47 &   0.85 &   0.44 &   0.74 &   1.48 &   1.95 &   0.97 & M3.5 \\
 G 171-59    &   0.60 &   0.61 &   0.82 &   0.55 &   0.78 &   1.28 &   1.95 &   1.42 & M1.5 \\
 G 171-61    &   0.65 &   0.62 &   0.80 &   0.59 &   0.80 &   1.18 &   1.96 &   0.97 & M1.5 \\
 G 132-4     &   0.47 &   0.51 &   0.85 &   0.47 &   0.76 &   1.42 &   1.95 &   0.95 & M3.5 \\
 LP 765-74   &   0.63 &   0.58 &   0.81 &   0.55 &   0.79 &   1.21 &   1.94 &   0.98 & M1.5 \\
 G 172-22    &   0.44 &   0.46 &   0.86 &   0.45 &   0.76 &   1.55 &   1.96 &   0.96 & M3.5 \\
 G 69-32     &   0.42 &   0.39 &   0.80 &   0.39 &   0.68 &   1.66 &   1.98 &   1.36 & M3.5 \\
\enddata
\end{deluxetable}

%% file: Reid.tab5.tex
\begin{deluxetable}{rrrrrrrrrc}
\tabletypesize{\scriptsize}
\tablecolumns{9}
\tablenum{5}
\tablewidth{0pt}
\tablecaption{Astrometry and photometry of spectroscopic targets}
\tablehead{\colhead{name} &
\colhead{ $\alpha$ (2000)} & \colhead{$\delta$} & \colhead{$m_r$} &
\colhead {J} &  \colhead {H} &\colhead {K$_S$} & \colhead{M$_J$} 
& \colhead{d (pc.)} }
\startdata
NLTT1  \\
LP 348-40$^{**}$ &  0:11:53.0  & 22:59:04    &    12.0 &   8.845 &   8.307 &   8.002 &   7.71 &   16.9$\pm$2.8 &    \\
 G 171-59            &  0:26:51.8  & 42:50:02    &    13.3 &   9.402 &   8.670 &   8.457 &   6.74 &   34.1$\pm$5.4 &    \\
 G 171-61            &  0:32:00.1  & 43:56:37    &    13.3 &   9.664 &   9.170 &   8.870 &   6.58 &   41.4$\pm$6.5 &    \\
 G 132-4$^{**}$    &  0:33:20.4  & 36:50:25    &    12.2 &   8.780 &   8.083 &   7.858 &   7.32 &   29.6$\pm$3.1 &    \\
 LP 765-74           &  0:37:36.3  & 15:11:52    &    12.3 &   9.087 &   8.432 &   8.236 &   6.75 &   29.3$\pm$4.6 &    \\
 G 172-22            &  0:46:29.9  & 50:38:39    &    13.6 &   9.968 &   9.329 &   9.082 &   7.54 &   30.6$\pm$4.8 &    \\
 G 69-32             &  0:54:48.0  & 27:31:03    &    14.1 &  10.325 &   9.734 &   9.444 &   8.34 &   24.9$\pm$4.5 &    \\
\enddata
\tablecomments{ \\
$^{**}$ identifies stars with formal distance estimates $d \le 20$ parsecs  \\
$^*$ identifies stars with formal distance estimates within $1\sigma$ of 20 parsecs  \\
1 - stars with JHK$_S$ from the 2MASS All-Sky data release.
}
\end{deluxetable}

%% file: Reid.tab6.tex
\begin{deluxetable}{rrrrrrrrrrrc}
\tabletypesize{\scriptsize}
\tablecolumns{12}
\tablenum{6}
\tablewidth{0pt}
\tablecaption{Early-type M dwarfs from the revised NLTT: Hipparcos stars}
\tablehead{\colhead{NLTT} &
\colhead{ Name} & \colhead{N$_{Hip}$} & \colhead{$\alpha$} &
\colhead {$\delta$} &  \colhead {V}  &\colhead {J} & \colhead{H} & 
\colhead{K$_S$} & \colhead {$\pi_{Hip}$} & \colhead {M$_J$} & \colhead {Comments}}
\startdata
    157 & Gl   3     &    436&  0:05:17.69 & -67:49:57.3 &   8.49&   6.492&   5.872&   5.737&  62.5$\pm$0.8&   5.47 &  \\
   1263 & Gl  18     &   1936&  0:24:25.93 & -27:01:36.4 &   7.92&   6.204&   5.700&   5.544&  55.5$\pm$1.1&   4.93 &  \\
   2941 & Gl  42     &   4148&  0:53:01.14 & -30:21:24.9 &   7.15&   5.537&   5.049&   4.889&  71.0$\pm$0.8&   4.79 & Note 1 \\
   4527 & Gl  57     &   6351&  1:21:34.59 & -41:39:23.1 &  10.15&   7.340&   6.754&   6.553&  59.3$\pm$1.6&   6.20 &  \\
   5563 & +14:251    &   7765&  1:39:56.06 &  15:15:33.7 &   8.68&   7.006&   6.454&   6.309&  61.3$\pm$36.7&   5.94 & not in CNS3, Note 2 \\
   7002 & LHS 1339   &   9786&  2:05:48.55 & -30:10:35.9 &  12.15&   8.389&   7.863&   7.558& 105.6$\pm$3.0&   8.51 &  \\
   7535 & LHS 1377   &  10617&  2:16:41.20 & -30:59:18.2 &  12.02&   7.975&   7.380&   7.092&  70.2$\pm$2.9&   7.21 & cpm LHS 1376 \\
   8126 & Gl 100AB   &  11565&  2:29:01.69 & -19:58:45.0 &   8.78&   6.586&   5.983&   5.817&  52.9$\pm$2.0&  5.5 & Note 3 \\
\enddata
\tablecomments{ V magnitudes from the rNLTT \\
1. J, H from final 2MASS data release. \\
2. BD+14:251: Optical companion affects Hipparcos astrometry; 
(V-K) colours give a photometric parallax of (m-M)$\sim$2.2 mag. \\
3. Unresolved in 2MASS scans - M$_J$ values estimated based on M$_V$ from CNS3 and average
main-sequence colours. \\
  Gl 100B - V=11.5, M$_J=6.7$; Gl 157B - V=11.8, M$_J=7.2$; Gl 200B - V=11.7, M$_J=6.7$; 
Gl 225.3B - V=8.3, M$_J=5.8$, Gl 225.3C - V=9.3, M$_J=6.5$; Gl 340B - V=8.1, M$_J=4.9$;
GJ 1181B - V=14.1, M$_J$=8.4; Gl 627B - V=7.85, M$_J$=4.9; GJ 1294B - V=10.6, M$_J$=6.8 \\ 
4. Gl 171.2A: Gl 171.2B is a DC8 white dwarf, M$_V = 15.8$, $\delta \sim 150$ arcseconds \\
5. Gl 225.2ABC: all three components merged on 2MASS - deconvolution assumes
average main-sequence colours. van Altena (1983)  measure $\pi = 0.058\pm0.0128$, but
the derived (V-K$_S$) colours suggest a distance of $\sim30$ parsecs. \\
6. Binary companion included in NLTT sample.}
\end{deluxetable}

%% file: Reid.tab7.tex
\begin{deluxetable}{rrrrrrrrrrc}
\tabletypesize{\scriptsize}
\tablecolumns{11}
\tablenum{7}
\tablewidth{0pt}
\tablecaption{Early-type M dwarfs from the rNLTT with no Hipparcos data}
\tablehead{\colhead{NLTT} &
\colhead{ Name} & \colhead{$\alpha$} & \colhead {$\delta$} &  \colhead {V$_{rNLTT}$}  
&\colhead {J} & \colhead{H} & 
\colhead{K$_S$} & \colhead {$\pi$ (mas)} & \colhead {M$_J$} & \colhead {Comments}}
\startdata
    629 & DV Psc     &  0:13:09.20&  5:35:43.0&  10.72&   8.040&   7.425&   7.262& 71$\pm$40 & 7.25/7.25 & Note 1  \\
   7251 & +45:565    &  2:11:43.26& 45:45:38.3&   9.83&   7.892&   7.433&   7.283& 24$\pm5$ & 4.8 & Note 2\\
  13091 & HD 27678   &  4:21:15.86&-20:11:26.1&   8.63&   6.845&   6.417&   6.287& 35$\pm$7 & 4.6 & Note 2\\
  17178 & HD 50841   &  6:56:16.50& 32:26:40.1&   8.64&   6.816&   6.382&   6.266& $33\pm$10 & 4.4 & Note 3\\
  18883 & G 111-51   &  8:05:01.85& 48:14:19.6&   9.31&   7.243&   6.729&   6.633& $34\pm7$ & 4.9 & Note 2 \\
  20631 & HD 76772   &  8:57:36.35&-19:08:02.2&   8.44&   7.133&   6.784&   6.691& 23$\pm$5 & 3.9 & Note 2\\
  25155 & HD 305544  & 10:42:47.27&-60:02:38.9&   9.18/9.6&   7.372&   6.987&   6.939& 16$\pm4$ & 4.1/4.4 & Notes 2, 4 \\
  25925 & HD 95174A  & 10:59:38.31& 25:26:15.5&   8.57&   6.444&   5.972&   5.842& 52$\pm9$ & 5.0 & Note 2\\
\enddata
\tablecomments{ 1. DV Psc: equal-magnitude eclipsing binary system  (Lu \etal, 2001). \\
2. Distance estimated using (M$_V$, (V-K$_S$)) relation from Paper I. \\
3. HD 50841: parallax from Yale parallax catalogue. \\
4. HD 305544: Yale parallax catalogue lists $\pi=10\pm12$ mas. 
}
\end{deluxetable}

%% file: Reid.tab8.tex
\begin{deluxetable}{rcrrrrcrc}
\tabletypesize{\scriptsize}
\tablecolumns{9}
\tablenum{8}
\tablewidth{0pt}
\tablecaption{Additional stars in the 2M2nd 20-parsec sample }
\tablehead{\colhead{Name } & \colhead {2MASS ID} & \colhead {V} &
\colhead{ J} & \colhead{H} & \colhead{K} & \colhead {Sp. Type} &  \colhead {(m-M)} & \colhead {Comments}}  
\startdata
 & PMSU non-NLTT & \\
  GR4         & 2M0015367-294600&14.32& 9.78$\pm$0.03& 9.25$\pm$0.02& 8.91$\pm$0.03& M3.5&  1.44$\pm$0.40 &\\
  GJ 1009     & 2M0021560-312421&11.16& 7.68$\pm$0.01& 7.05$\pm$0.03& 6.77$\pm$0.01& M1.5&  1.30$\pm$0.09 &\\
  GJ 1033     & 2M0113239-225407&14.16& 9.91$\pm$0.02& 9.30$\pm$0.02& 9.04$\pm$0.03& M4.0&  1.44$\pm$0.15 &\\
  Hy207       & 2M0330311+200555&10.79& 8.05$\pm$0.02& 7.40$\pm$0.01& 7.43$\pm$0.03& K7.0&  1.36$\pm$0.27 & \\ 
  Steph 430   & 2M0354256-090931&11.23& 7.84$\pm$0.02& 7.18$\pm$0.03& 6.98$\pm$0.02& M1.0&  1.37$\pm$0.40 &\\
  Steph 497   & 2M0437374-022928&10.59& 7.30$\pm$0.01& 6.64$\pm$0.05& 6.39$\pm$0.01& M0.5&  1.09$\pm$0.40 &\\
  Steph 546A  & 2M0506494-213503&10.28& 7.03$\pm$0.01& 6.41$\pm$0.01& 6.11$\pm$0.02& M1.5&  0.36$\pm$0.40 &\\
\enddata
\end{deluxetable}

%% file: Reid.tab9.tex
\begin{deluxetable}{rrrrrc}
\tabletypesize{\scriptsize}
\tablecolumns{6}
\tablenum{9}
\tablewidth{0pt}
\tablecaption{A rough guide to absolute J magnitudes for K/M dwarfs}
\tablehead{\colhead{M$_J$} & \colhead { (V-J)} & \colhead { (J-K$_S$) } &
\colhead{ (H-K$_S$)} & \colhead{M$_V$} & \colhead {Sp. Type} } 
\startdata
4.5 & 1.4 & 0.55 & 0.11 & 5.9 & K2 \\
5.5 & 2.3 & 0.8 & 0.16 &7.8 & M0 \\
6.5 & 3.3 & 0.8 & 0.20 &9.8 & M1 \\
7.5 & 4.0 & 0.75 & 0.25 &11.5 & M3 \\
8.5 & 4.4 & 0.8 & 0.25 &12.9 & M4 \\
9.5 & 4.6 & 0.85 & 0.30 &14.1 & M5 \\
10.5 & 6.2 & 1.0 & 0.35 &16.7 & M6 \\
11.5 & 7.5 & 1.25 & 0.42 &19.0 & M8 \\
12.5 & 8.5 & 1.5 & 0.55 &21.0 & L0 \\
\enddata
\end{deluxetable}

%% file: Reid.apptab1.tex
\begin{deluxetable}{rrcl}
\tabletypesize{\scriptsize}
\tablecolumns{4}
\tablenum{A1}
\tablewidth{0pt}
\tablecaption{Photometric outliers in the rNLTT}
\tablehead{\colhead {N$_{NLTT}$} & \colhead{Name} 
& \colhead{ m$_r$} & \colhead{ Comment} }
\startdata
   166$^*$&   LP 824-355&  14.4&  cpm LP 824-355, m$_r$=13.2, sep=7", V mag. probably affected \\
  1574&   LP 240-22 &  18.8&  mismatch to 2MASS data for cpm, LP 240-23, 18.1 \\
  3411  & G 268-98   & 13.5&  mismatch to 2MASS data for cpm, BD-16:168, 9.0                                 \\
  5244&   LP 468-114&  17.0 & mismatch to 2MASS data for cpm, LP 468-115, 13.2 \\
  6886&   LP 829-5  &  18.8&  WD?, mismatch to 2MASS data for cpm, LP 829-4, 15.2 \\
  7084$^*$&   G 73-27   &  14.3&  cpm LP 469-118, 14.4, sep=3", JHK affected in 2M2nd \\
  7679&   LP 245-18 &  17.0&  WD, mismatch to 2MASS data for cpm, G74-15 = GJ 3151A \\
  8721&   LP 470-44 &  18.7&  mismatch to 2MASS data for cpm, LP 470-43, 13.1 \\
\enddata
\tablecomments{$^*$ - additional observations included in the current program} 
\end{deluxetable}

%% file: Reid.apptab2.tex
\begin{deluxetable}{rrcl}
\tabletypesize{\scriptsize}
\tablecolumns{4}
\tablenum{A2}
\tablewidth{0pt}
\tablecaption{Comments on rNLTT stars}
\tablehead{\colhead {N$_{NLTT}$} & \colhead{Name} 
& \colhead{ m$_r$} & \colhead{ Comment} }
\startdata
  2804  & LP 350-19  & 12.0&  merged with cpm, LP 350-20, 11.7                                               \\
  2805  & LP 350-20  & 11.7&  merged with cpm, LP 350-19, 12.0                                               \\
  3462  & LP 406-76  & 11.7&  merged with cpm, LP 406-77, 11.7                                               \\
  3463  & LP 406-77  & 11.7&  merged with cpm, LP 406-76, 11.3                                               \\
  3482  & LP 194-20  & 13.7&  mismatch to 2MASS data for cpm, G 132-51, 12.7                                 \\
  3642  & LP 294-49  & 14.1&  merged with cpm, G 132-54, 14.9                                                \\
  3916  & LP 767-16  & 13.6&  merged with cpm, LP 767-17, 14.1                                               \\
  3917  & LP 767-17  & 14.1&  merged with cpm, LP 767-16, 13.6                                               \\
\enddata
\end{deluxetable}